\documentclass[11pt,twoside]{article}
\usepackage{cozumel2005}
\usepackage{epsf}
\usepackage{psfig}
\usepackage{lscape}
\usepackage{graphicx}
\begin{document}
\title{Revisiting and assessing uncertainties in stellar populations
synthesis models}    
\author{M. Cervi\~no and V. Luridiana}   
\affil{Instituto de Astrof\'\i sica de Andaluc\'\i a (IAA-CSIC), Camino bajo de Huetor 50, 18008 Granada, Spain}    

\begin{abstract} 
In this review we address the uncertainties implicit in evolutionary synthesis model computations. After describing
the general structure of synthesis codes, we discuss several source of uncertainties that may affect their results.
In particular, we discuss the uncertainties arising in the computation of isochrones from evolutionary tracks;
those related to atmosphere models; those that are a consequence of the incompleteness
of the input ingredients; and those associated with the computational aspect used in synthesis codes.
We also discuss the issue of the inclusion of distributed properties in synthesis models,
an issue that will become relevant in the next future; as a paradigm of this case, we illustrate
the difficulties implied by the inclusion of tracks with rotation in synthesis models.
Finally, we describe several examples of the statistical approach to population synthesis.

We report on the failure of the fuel consumption theorem (FCT) and the isochrone synthesis code
to produce mutually consistent results. However, we argue that FCT and isochrone synthesis results 
are reliable for application to real systems in the wavelength range where they coincide. 

On the constructive side, we derive several useful survival strategies to bypass uncertainties.
We show that single stellar populations at the turn-off ages of the tabulated tracks can be
safely compared, as they are scarcely affected by the interpolation scheme
used to compute isochrones.
Finally, we suggest to use derivative quantities, such as the SN-rate, as bug detectors.

On the recommendation side,
we advocate for greater transparency and more documentation in synthesis modeling.
 We also ask stellar model makers to think of us and include more mass values in the tracks.

\end{abstract}

\section{Introduction}                

Synthesis codes are used for a variety of scopes.
Their goals range from testing the accuracy of evolutionary tracks to explaining the
physical characteristics of distant galaxies, to measuring the extinction of outer galaxies,
or to computing suitable inputs to photoionization models.
They are a natural link between our knowledge of individual stars and the properties of stellar ensembles,
and as such they are an invaluable tool in the analysis of stellar populations.

Given this flexibility, it is tempting to forget their limitations;
yet an awareness of the uncertainties affecting 
the results  of synthesis codes is necessary if one wants to take full advantage of their power.
Limitations arise from different issues.
One is specialization: although, as a generic tool, synthesis codes can be used to address a variety of problems, 
each code is specialized in a particular niche of the general physical problem. 
Stretching the result of a particular code out of its natural boundaries is a frequent mistake, 
and statements like "Everybody knows that the results of synthesis codes are code-dependent" are unfortunate outcomes
of code misuse.
Other limitations are a direct consequence of the incompleteness of the ingredients used;
still others depend on the features of the numerical methods employed by the code.
All these uncertainties must be known and understood whenever synthesis models are used.

In order to assess the uncertainties in stellar population synthesis, it is mandatory 
to understand what evolutionary synthesis codes do, which their ingredients and their intrinsic 
assumptions are, how such assumptions are managed by the codes, and to which 
situations the results of the codes can be applied. Following this scheme, the structure of this 
review is the following. We will first describe the general structure of any synthesis code in section \S\ref{sect:general} 
In section \S\ref{sect:ingredients} we will discuss the uncertainties and limitations of the ingredients and their impact on 
the interpretation of the results; we include here an extensive discussion of how isochrones can be obtained from evolutionary tracks, and their associated uncertainties. In section \S\ref{sect:numerical} we will discuss the problems related to numerical methods. Section \S\ref{sect:future} is devoted to the challenges posed by the inclusion of stellar rotation and variability in synthesis codes. 
In section  \S\ref{sect:discussion} we will discuss the interpretation of results and in section  \S\ref{sect:conclusions} we will draw the conclusions.

Before starting, let us apologize to those who expect to find an extensive set of references in this review. Most of the issues addressed here are also discussed by other papers reporting on the results of particular evolutionary synthesis codes, sometimes described at length and sometimes mentioned in just a few sentences. Instead of citing all these papers, we have chosen to maintain the number of references to a minimum preferring to discuss exhaustively the issues related to the uncertainties in models. For more traditional reviews on synthesis models we refer to \cite{Mar03} and \cite{Pop05}. We also refer to the papers by \cite{CWB96,dG03,And04,dG05}, which discuss some issues that are not addressed here.

\section{Overview of the problem}\label{sect:general}

The general problem addressed by synthesis codes is the computation
of the luminosity
$L_{\mathrm{tot}}$ emitted by an ensemble of 
$N_{\mathrm{tot}}$ stars -- a stellar population.
From a theoretical point of view, this problem 
can be characterized in the following basic ways. 
 
If the luminosities $\ell_i^*$ of the individual stars are known, 
the total luminosity $L_{\mathrm{tot}}$ 
is obtained trivially as the sum of all the $\ell_i^*$ values:

\begin{equation}
L_{\mathrm{tot}} = \sum_{i=1}^{N{_\mathrm{tot}}}  \, \ell_i^*.
\label{eq:Ltot_sumindv}
\end{equation}

\noindent This circumstance is not very frequent; its most common 
examples are Monte Carlo synthetic clusters in the theoretical domain,
and resolved stellar populations in the observational one.

In the more usual situation in which the luminosities of individual 
objects are not known, a different approach must be adopted.
In this case, it can be assumed that the emission of the ensemble is given by the sum of the 
contributions of $N_{\mathrm{class}}$
different classes of stars:

\begin{equation}
L_{\mathrm{tot}} = \sum_{i=1}^{N{_\mathrm{class}}}  \,w_i \ell_i,
\label{eq:Ltot_sum}
\end{equation}

\noindent each including a relative number of stars $w_i$ with average luminosity $\ell_i$.
A class is formed by fairly homogeneous stars,
a group of contiguous classes can be identified with a stellar evolutionary phase\footnote{Here we adopt the definition of stellar evolutionary phase used
in stellar evolutionary theory, i.e. a particular stage of stellar evolution characterized by well-defined structural features, 
usually a given mode of nuclear burning: examples of evolutionary phases are the main sequence and the red giant branch. 
We warn, however, that the expression 'stellar evolutionary phase' is sometimes
used more loosely in population synthesis literature, indicating any of the arbitrarily defined classes of Eq.~\ref{eq:Ltot_sum}.},
and the whole of classes
represents a particular ensemble of stars.

As will be explained in the following, the coefficients $w_i$ can be computed
with the results of stellar evolution and stellar atmospheres theories
weighted by the stellar birth-rate,
a function returning the number of stars of each given initial mass 
born at a given time.

Stellar evolution theory describes the time evolution of model stars. Its
results are quantities in the theoretical space, 
e.g. the bolometric luminosity $L_\mathrm{bol}$, 
the effective temperature $T_\mathrm{eff}$, and so on. However, the results must be compared with 
observable quantities, such as luminosities in a given band or spectral indices.  Hence, a 
transformation from the theoretical space to the observational one is required: this is the domain of the theory of stellar atmospheres. 
Such transformation requires using  a collection of templates of the emission of the considered source (i.e. an atmosphere library). 
In mathematical terms, this corresponds to defining the observable quantity $\ell_i$ 
in terms of quantities in the theoretical space, i.e. $\ell_i(L_\mathrm{bol}, T_\mathrm{eff},...)$.

The stellar birth rate tells us how many stars with a given initial mass are born in the cluster at a given time.
It is often assumed \citep[e.g.,][]{TG76} that the stellar birth-rate can be separated into two functions independent 
of each other: the Star Formation History (SFH), which gives the number
of stars born in a given time, and the Initial Mass Function (IMF), which
gives the relative number of stars born as a function of the mass. 
This assumption, which will be discussed briefly in \S\ref{sect:birthrate}, is the strongest {\it ad hoc} assumption of current synthesis
codes \citep{Shore03}.

Stellar evolution theory, stellar atmospheres theory and the stellar birth rate are the basic input ingredients
of evolutionary synthesis codes, which are tools designed to solve Eq.~\ref{eq:Ltot_sum}.
From the point of view of synthesis codes, these input ingredients can be classified in two categories: input data and input parameters. 
Stellar tracks and model atmospheres, which are often included as built-in libraries, are the input data. 
The stellar birth rate and a few other quantities, such as the lower and upper mass limits of the IMF, the age and the metallicity, 
are the input parameters, that is they are switches that tell the code how to combine the input data to produce a model population.
Usually, the input data available to a given code are selected and controlled by the author of the code. 

\begin{figure}[!ht]
\includegraphics[width=\textwidth]{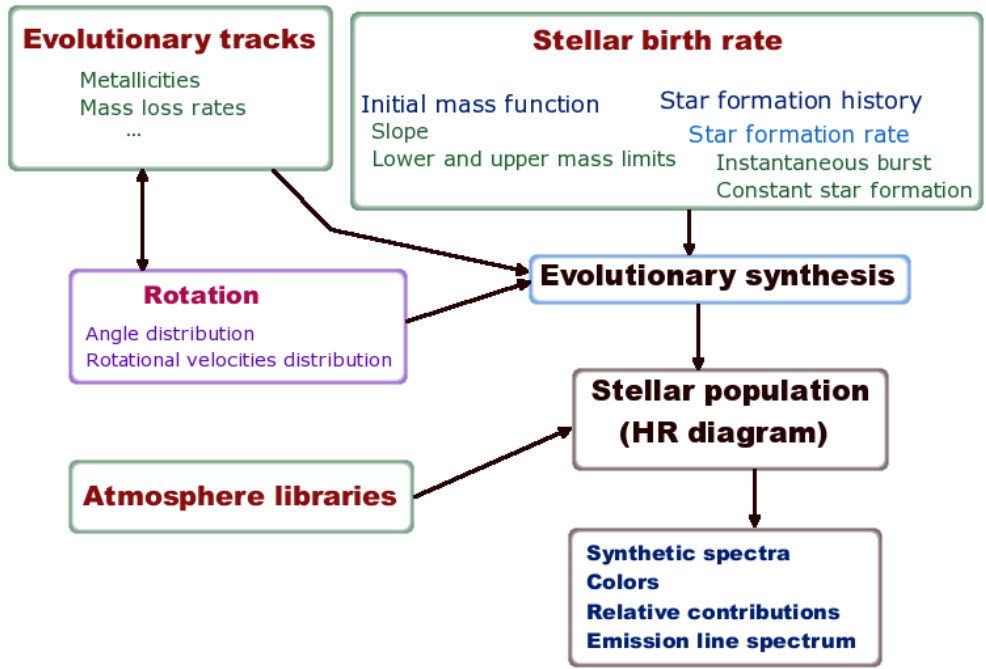}
 \caption{General structure of evolutionary synthesis codes.}  
 \label{fig:models}
\end{figure}

The general structure of synthesis codes is summarized in Fig. \ref{fig:models}. 
Input ingredients will be described in the next section.

\subsection{Evolutionary tracks and isochrones}\label{sect:tracks}

The stellar models entering a synthesis code are made up of two different parts: evolutionary tracks and stellar atmospheres templates. The former describe the inner structure and evolution of stars, and the latter their outer, radiation-emitting layers. The computation of these two parts require completely different physics and is carried out by different codes: hence, evolutionary tracks and model atmospheres or empirical libraries are provided separately and then coupled together in the synthesis code. 

Evolutionary tracks are so called because they carry information on the path ({\it track}) that a star follows on the HR diagram throughout its lifetime ({\it evolutionary}). This naming convention is somewhat reductive, since evolutionary tracks in fact include much more information than the one displayed in the HR diagram: in fact, each time-point of an evolutionary track also contains details of the stellar structure, that is the dependence on radius of variables such as temperature, density, chemical composition, nuclear energy production rate, etc. Therefore, they are in fact models of stellar structure and evolution: here, however, we will follow the customary use and refer to them as evolutionary (or stellar) tracks. 
At first order, an evolutionary track is identified by its initial mass and metallicity. 

The tracks determine strongly the scope and the limitations of a synthesis model, so the choice of which tracks to use is an important one both for code developers and for end users. In the following, we will define a set of stellar tracks as a batch of stellar tracks of different initial masses computed under homogenous assumptions. Sets of stellar tracks available in the literature differ from each other in many respects:  the metallicity values available,
 the mass range covered, the treatment of mass loss, the inclusion of rotation,
plus other lesser-order effects such as the treatment of convection and the opacity tables included. Stellar track makers are often specialized in a small niche of stellar evolution theory, e.g. low-mass or massive stars. For this reason, the choice of the stellar tracks to be used in a synthesis model depends a lot on the kind of model one wants to build. For example, a synthesis model of an old population requires accurate tracks of low-mass stars, while a model of a young population requires accurate tracks of massive stars. 

Ideally, stellar tracks with any desired value of the stellar parameters should be available. Unfortunately, since evolutionary tracks are costly to compute and store, available models have only a very limited subset of all the possible metallicity and initial mass values. Typically,  a given author usually provides sets of tracks corresponding to less than ten metallicity values to cover a range of several orders of magnitude. A second issue related to evolutionary tracks is that a given set of tracks does not necessarily cover all the evolutionary phases. In such a case the synthesis model maker must choose between keeping a self-consistent set of (incomplete) tracks or mixing different sets of tracks (sometimes with different physical assumptions). 
Therefore, when using or computing a synthesis model, one should be aware of the specific features of the stellar models included in it. 

Once the set of evolutionary tracks has been defined, one has a set of functions that describe the evolution as a function of time $t$ of stars with  given initial mass $m$, $f_\mathrm{track}(t;m)$. The next step is transforming it into an isochrone. 
An isochrone is defined by the population of coeval stars of different masses at a given age. Isochrones can be represented in different plans: as example, an isochrone in the theoretical HR diagram is a parametric curve $f_\mathrm{iso}(m;t) = [L_\mathrm{bol}(m;t), T_\mathrm{eff}(m;t)]$ ($m$ being the initial mass). 

Isochrones are like snapshots of evolving populations. Conceptually, they are an intermediate step in the transition from information on individual stars to information on composite populations. Not all codes perform this step, as isochrones are available in the literature so they can directly be used in a code, bypassing the need for evolutionary tracks. But those codes that use tracks as input must interpolate between evolutionary tracks to obtain the isochrones at a given value of $t$.

\subsection{Atmosphere libraries}\label{sect:atmospheres}

The atmosphere is the outermost part of a star. The mass of a star's atmosphere is negligible when compared to the star's total mass, so it has no effect on the stellar structure and evolution; but, by definition, its optical depth at most of frequencies is small, so it is the region where the emitted spectrum forms. 

A synthesis code needs an atmosphere library to transform the theoretical
quantities obtained with the evolutionary tracks into observational
ones. There are two possible alternatives:

\begin{enumerate}

\item Grids of theoretical atmospheres, composed of
atmosphere models. The main challenge of the models is that 
in the atmosphere the radiation field is strongly anisotropic, so the transfer equation must be solved explicitly. Different approximations can be adopted to solve this problem in model atmospheres, and one must be aware of the implications of choosing among different models, as different authors make different assumptions when they face the extraordinarily complex problem of computing a stellar atmosphere. 
Model atmospheres are generally given as a function of metallicity, gravity ($g$) and effective temperature. 

\item Grids of empirical atmospheres, composed of observed
stars. These stars must be calibrated in flux and their
physical properties must be well determined. Since the observed stars
are used as templates for the stellar classes that have a bolometric luminosity different from the one of  the observed star, a normalization it is also
necessary.

\end{enumerate}

Analogously to what happens with evolutionary tracks, the available models cover only a very limited subset of all the parameter space.
For a more detailed paper about the atmospheres libraries used
in synthesis codes, we refer to the contribution by Gustavo Bruzual in these
proceedings.

\subsection{Stellar birth rate}\label{sect:birthrate}

In order to obtain the weight $w_i$ of each stellar class, we need a recipe to compute the number of stars 
populating each class at any given time. 
To this scope, we need to know which kind of stars occupy a certain position at the time considered, and how many of them there are.
The first piece of information is given by evolutionary tracks combined with model atmospheres, which we have introduced in \S\ref{sect:tracks} and \S\ref{sect:atmospheres}.
The second piece of information is the stellar birth-rate. In order to simplify the subject, we will assume in the following that the mass and the time dependence of the stellar birth rate can be separated, as most synthesis codes also assume. 
This is a very strong hypothesis, which is assumed {\it ad hoc} 
to simplify the handling of the equations, and which is not necessarily a realistic one:
for example, it is trivially false if
stellar winds from massive stars inhibit the formation of low mass stars,
or in any other case in which the IMF keeps memory of previous star
formation episodes. See \cite{gtt05, Shore87,SF95} for more details on
this topic.

For simplicity, let us also assume that the SFH is described by a Dirac delta function (the star-formation episode is extremely concentrated in time). This scenario has been labeled in different ways as coeval star formation, instantaneous burst or simple stellar population (SSP). This mathematical approximation 
allows us to neglect the chemical evolution of stars in the cluster (since all the stars are formed with the same
initial metallicity) and to define a zero-point in the timeline.

To compute the expected emission of a stellar population in the SSP hypothesis, 
it is only necessary to provide its {\it initial} mass composition, which is defined by the IMF. The concept of IMF is rooted in measurements of the relative frequencies of stars with different masses, which are observed to be fairly constant across different stellar populations. The IMF is often approximated either by a power-law or by a sum of power-laws over different subranges: 

\begin{equation}
\varphi_\mathrm{M}(m) = \frac{dN}{dm} \propto m^{-\alpha}.
\end{equation} 

\noindent A typical value for the power-law exponent is the one by Salpeter, $\alpha=2.35$  \citep{Sal55}. Alternative IMFs have also been proposed \cite[e.g.][]{MS79,Rana87,Kro01}. The IMF in the low-mass tail is poorly known, due to incomplete detection. The IMF in the high-mass tail is poorly known, due to the intrinsically low mass counts. We refer to these proceedings for more extensive reviews on the subject.

\subsection{Types of synthesis codes}\label{sect:types}

With these ingredients it is now possible to compute the contribution of the different stellar classes: since each stellar class present at a given time can be defined by a range of initial stellar masses, its contribution, $w_i$,  is simply given by:

\begin{equation}
w_i = \int_{m_i^{\mathrm{up}}}^{m_{i}^{\mathrm{low}}} \varphi_\mathrm{M}(m) \, dm,
\label{eq:wi}
\end{equation}

\noindent where $m_i^\mathrm{low}$ and $m_{i}^{\mathrm{up}}$ 
are the lower and upper limits of the $i$-th mass bin that gives rise to the $i$-th stellar class (the specific way in which these limits are defined varies from code to code). The codes that compute the contributions of the different stellar classes in this way, based on mass bins, are usually called 
{\it isochrone synthesis codes}. 

\begin{figure}[!ht]
\includegraphics[width=\textwidth]{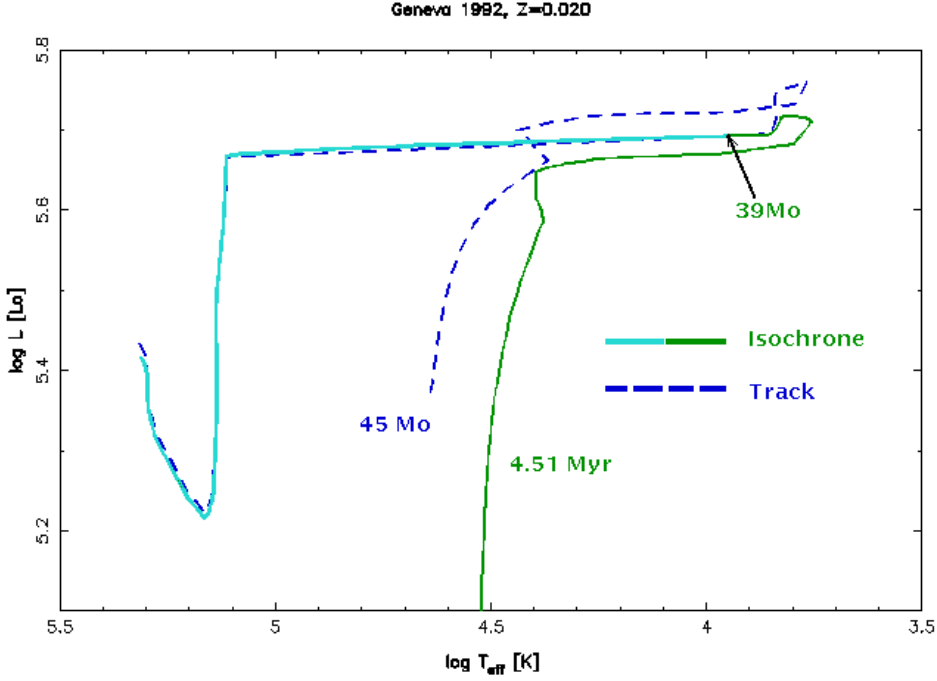}
\caption{Isochrone at 4.51 Ma and evolutionary track of 45 M$_\odot$ taken from \cite{Schetal92}.
The mass point where the isochrone is equivalent to the track, 39 M$_\odot$, is marked in the diagram.} 
\label{fig:FCT}
\end{figure}

There is at least another way to compute these contributions, which is based on
the evolutionary times of the different phases.
To understand this point,
let's consider how an isochrone of a given age is populated.
The post-MS portion of an isochrone is populated by all those stars who have already evolved out of the MS
but that have not died yet. Since post-MS evolutionary times are much shorter than MS times, 
the difference in mass between the turn-off (TO) star (the star that is just about to leave the MS)
and the most evolved star is comparatively small.
It can be shown that the older the SSP, the smaller such difference:
asymptotically, all the post-MS isochrone portion is populated by stars with the TO mass,
and the isochrone tend to converge to the corresponding evolutionary track to the degree that, from the TO on, they 
merge completely. 
This is shown with an example in Fig. \ref{fig:FCT}, which 
illustrates the point in a less extreme situation (i.e. one in which the merging point takes place at a mass point located beyond the TO).

The consequence of the merging of isochrones and evolutionary tracks for late evolutionary stages is that
the post-MS luminosity of a SSP population is proportional to the fuel available to the TO star.
Therefore, the contribution $w_i$ of a given post-MS stellar classe can by expressed by Eq. \ref{eq:wiFCT}:

\begin{equation}
w_i = \int_{m_i^{\mathrm{up}}}^{m_{i}^{\mathrm{low}}} \varphi_\mathrm{M}(m) \, dm = \varphi_\mathrm{M}(m_\mathrm{eq}) \, \delta m_i = \varphi_\mathrm{M}(m_\mathrm{eq}) \, \left| {\frac{d m_\mathrm{eq}}{d t}} \right| \, \delta t_i,
\label{eq:wiFCT}
\end{equation}

\noindent where $m_\mathrm{eq}$ is the mass point in the isochrone where the track and the isochrone become equivalent, 
and $\delta t_i$ is the duration of the evolutionary segment corresponding to the $i$-th stellar class.
This way of computing the post-MS contribution to luminosity is known as the {\it fuel consumption theorem} \citep[FCT;][]{RB86,Buzz89}. 
We refer to these papers for further details.
Synthesis codes based on Eq.~\ref{eq:wiFCT} are usually called {\it fuel consumption codes}. 

In the practice, the distinction between isochrone synthesis and fuel consumption codes is not a sharp one,
as most codes make use of both methods. For example, since the FCT makes explicit reference to post-MS phases,
FCT codes compute the contribution of the MS as isochrone synthesis codes do. 
On the other hand, some isochrone codes make use of the FCT to include post-Asymptotic Giant Branch (post-AGB) evolution.

\section{Overview of the uncertainties}\label{sect:ingredients}

We can now move on to discuss the main sources of uncertainty in synthesis codes.
These can be grouped as follows:

\begin{enumerate}
\item The uncertainties in the evolutionary tracks used and in the transformation from tracks
to isochrones.
\item The uncertainties in the assumed stellar birth-rate.
\item The uncertainties in the assumed atmosphere libraries and their assignation to theoretical quantities.
\item The incompleteness of input ingredients.
\item The numerical methods used.
\end{enumerate}

These uncertainties will be discussed in the following. Throughout this discussion, 
it will be useful to keep as a reference the sketch of
the general structure of synthesis codes shown in Fig. \ref{fig:models}. 

\subsection{Uncertainties related to the evolutionary tracks and isochrones}\label{sect:uncertainty_tracks} 

\subsubsection{The homology assumption in track interpolations}\label{homology}

Some track providers (but not all) tabulate their tracks following a
sequence of {\it equivalent evolutionary points}. These are defined as
points in stellar tracks of different initial masses that can be related
by means of homology relations: i.e. the structures of
two models star of different initial masses are homologous at corresponding
equivalent evolutionary points \citep{KW90,Mow98}.
For exampe, in the case of a completely radiative main sequence (MS) star with a Kramer's
opacity law the homology relations are:

\begin{eqnarray}
r(m) &\propto & (\epsilon_0\kappa_0)^{1/20} \mu^{13/20} m^{4/5},\nonumber
\\
\ell_\mathrm{bol}(m) &\propto & \kappa_0^{-1} \mu^4 m^{3}, {\mathrm and}\nonumber \\
\tau_\mathrm{ms}(m) &\propto & \kappa_0  \mu^{-4} X  m^{2},
\label{eq:homo}
\end{eqnarray}

\noindent where $\kappa_0$ is the opacity, $\mu$ the mean molecular
weight, $\epsilon_0$ the energy production rate and $X$ the fraction of
hydrogen in the stellar core. These relations, together with the Stefan-Boltzmann
law,

\begin{equation}
\ell_\mathrm{bol} \propto r(m)^2 T_\mathrm{eff}(m)^4,
\label{eq:Lbol}
\end{equation}

\noindent allow to perform interpolations among tracks in the MS, assuming that the mass is constant throughout the evolution, thus obtaining tracks that are not tabulated. These relations provide
an interpolation scheme to obtain $\ell_\mathrm{bol}(m)$ and
$T_\mathrm{eff}(m)$ for any desired mass.

In fact, homology relations are a good approximation only for MS stars \citep{KW90}: 
beyond the MS the homology assumption always breaks down. However, in isochrone computations the interpolation scheme implicitly
assumes homology even beyond the MS. 

A second point to be addressed is that due to the presence of mass loss the 
assumed homology relations are not valid, since mass loss is  coupled 
with the stellar evolution in a non-trivial way that is not considered in the usual homology relations;
but, again, the interpolation scheme among tracks is maintained even if tracks with mass loss are used.

Summing up, the interpolation scheme which is customarily used to calculate tracks that are not tabulated
is based on an assumption which is, most of the times, not valid.
There are, however, some ages at which the homology assumption has only a minor impact on the computed models. 
These are the ages at which the stars with tabulated tracks are at the TO. As we have shown before,
post-MS stars have masses similar to the TO mass;
if the TO mass is a tabulated one, the contribution of post-MS stars will not be much affected by interpolation errors.
This observation permits to formulate a criterion for a reliable comparison among synthesis models:

\begin{center}
{\it  The most reliable ages for model use or comparison are
the MS turn-off ages of the input tracks.}
\end{center}

If differences appear at such ages, they reflect differences in the numerical methods used in each code, which must be further investigated.
This criterion is only valid, of course, if the codes have the same input ingredients.

\subsubsection{Fast evolutionary phases}\label{sect:fastphases}

\begin{figure}[!ht]
\includegraphics[width=\textwidth]{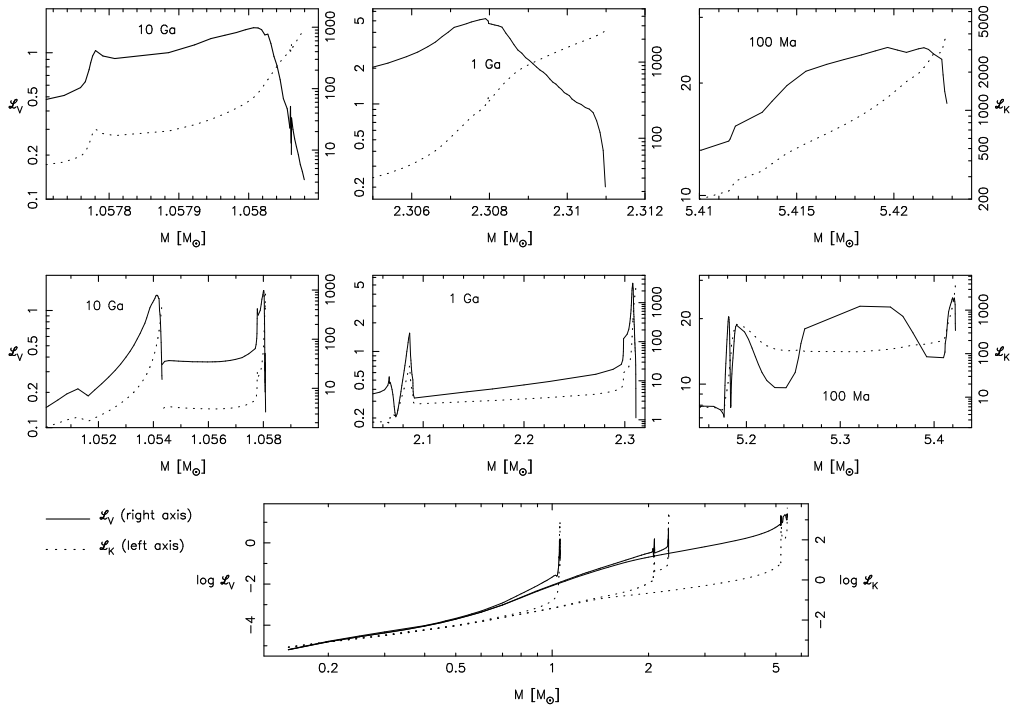}
\caption{Details of fast evolutionary phases in the V (solid line, left axis) and K (dotted-line, right axis) bands. Bottom panel: complete isochrones. Middle panels: blow-up of the mass axis, showing details of fast evolutionary phases at three different ages. Top panels: same as middle panels, with a more extreme blow-up. The isochrones have been computed by \cite{Gi02} from the evolutionary tracks by \cite{MG01}}
 \label{fig:iso}
\end{figure}

In post MS phases, luminosity is not always a well-behaved function of the mass, so  the isochrone is not always defined:
a typical isochrone features shallow sections as well as peaks and discontinuities in the $m$,$\ell$ plane, as shown in Fig. \ref{fig:iso}. Shallow sections correspond to quiescent phases of stellar evolution, where evolution is slow (e.g. the MS); peaks correspond to faster phases (e.g. the asymptotic giant branch); and discontinuities correspond to abrupt transitions between phases (e.g. the onset of central helium burning in intermediate mass stars) or jumps in stellar behavior (e.g., the transition between Wolf-Rayet [WR] and non-WR-type structures). Peaks and discontinuities will generically be labeled in the following as fast evolutionary phases.
It is interesting to realize that the expression {\it fast evolutionary phases} makes explicit reference to time variation, although by definition time is a constant throughout an isochrone! In fact, the term {\it fast} refers  to the behaviour of the star whose evolutionary track coincides with the isochrone in the post-MS phases. 
In terms of isochrones, the expression refers to regions where the derivative ${d\ell}/{dm}$ diverges.

Fast evolutionary phases are difficult to handle, because a small difference in the initial stellar mass yields a large difference (or an indetermination) in luminosity, so that the result  depends crucially on which luminosity is chosen to be representative of a given stellar class. In principle, this difficulty can be dealt with by defining stellar classes so that they are characterized by small luminosity ranges and they do not go across discontinuities; however, the available resolution in mass is governed by the evolutionary tracks used by the code, and is generally too low to resolve adequately such phases in synthesis models. This is a severe problem, since fast evolutionary phases are ubiquitous in post MS evolution, and at certain frequencies they bear a major weight in the luminosity balance. 

This problem has been present in synthesis codes since its very beginning \citep{TG76}.
In the particular case of discontinuities, 
the way in which it is tackled is often labeled a `technical
detail' of the computation and dismissed as unimportant,
and thus the papers describing evolutionary synthesis models
do not generally make any reference to its solution -- in spite of
its difficulty and of the potentially disastrous consequences of
incorrect assumptions.  Here are a few examples of the ways in which the
problem of discontinuities has been approached: a) in the Starburst99 synthesis
code \citep{SB99}, for certain metallicity values,  
an undocumented stellar track at 1.701 M$_\odot$ 
is added to the tabulated track of 1.70 M$_\odot$ by
\cite{Schetal92} and \cite{Schetal93}, to avoid stellar classes
going across the discontinuity of the stellar models' behavior
at such mass (C. Leitherer, D. Schaerer, \& G.
Meynet, private communication); b) to deal with the same problem, 
additional evolutionary tracks around the same mass range 
are used in the computation of the isochrones
by \cite{Casetal03} and \cite{Caretal04} (S. Degl'Innocenti, private communication)
and \cite{Broetal99} (E. Brocato, private communication); 
c) to avoid the intrinsic discontinuity in the isochrones, 
the same mass is used twice in the isochrones by \cite{Gi02}, 
namely at the end of the  red giant branch (RGB) and
at the beginning of the horizontal branch 
(S. Bressan, private communication).

\begin{figure}[!ht]
\includegraphics[width=\textwidth]{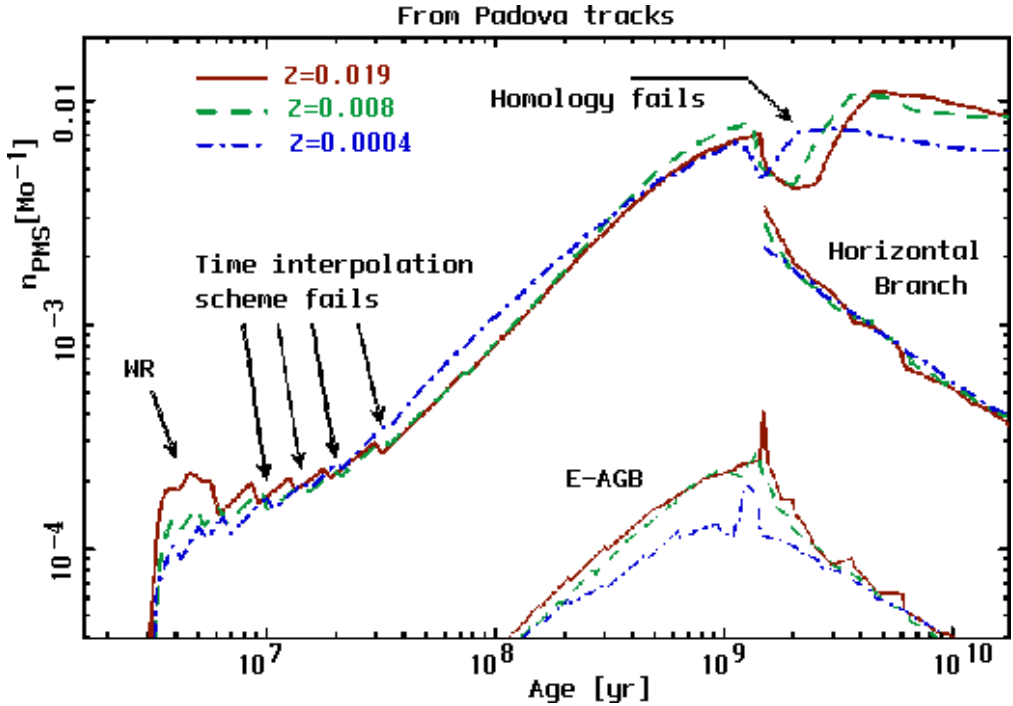}
\caption{Number of post MS stars as a function of time for different metallicities, obtained from the isochrones compiled by \cite{Gi02}.}
 \label{fig:homo}
\end{figure}

This problem shows up in the results of synthesis codes. For example,
Fig. \ref{fig:homo}, in which the number of post-MS stars is plotted as a function of time, 
shows a gap at ages around 10$^9$ years; the gap corresponds to the MS evolutionary time of stars 
with initial masses near the 1.7 M$_\odot$ discontinuity. 

The figure also shows the presence of small peaks  at young ages. The maximum of each peak indeed corresponds to the MS lifetime of the tabulated tracks. The peaks appear as a consequence of an incorrect time interpolation scheme and can be corrected by adopting a different one. This problem will be discussed in the next paragraph.

\subsubsection{Evolutionary times and time interpolations}\label{sect:time}

\begin{figure}[!ht]
\includegraphics[width=\textwidth]{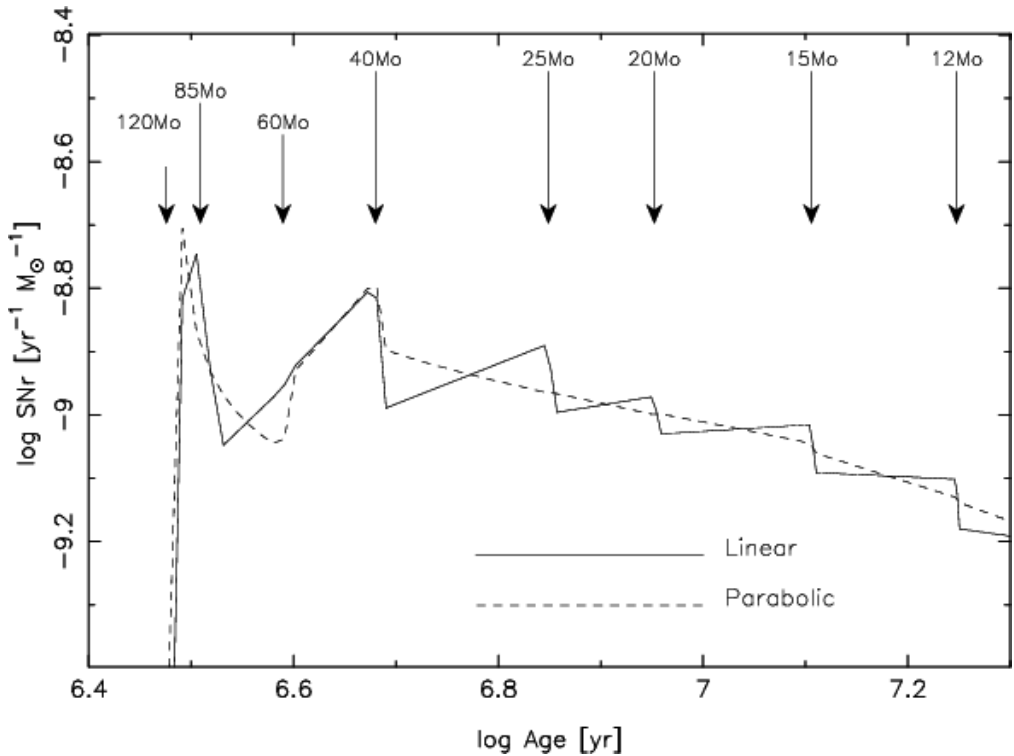}
\caption{SN rate vs. time using different interpolation techniques. The solid line corresponds to a linear interpolation in the $\log m - \log t$ plane. The short-dashed line corresponds to a parabolic interpolation. The evolutionary tracks used have been computed by \cite{Schetal92}: Star formation law: instantaneous; solid line, $\alpha = 2.35$, M$_\mathrm{up}$ = 120 M$_\odot$; Z=0.020. Figure taken from \cite{Cetal01}.}
\label{fig:SN}
\end{figure}

A further problem related to the computation of isochrones is the duration of the different evolutionary phases that are included in the tracks. 
In \S\ref{sect:tracks} we have shown that, in order to obtain isochrones, it is necessary to interpolate among the tabulated tracks so that tracks with intermediate values are obtained.
After that, it is necessary to interpolate in time among different tracks to obtain the isochrone at a given age. 
The usual scheme assumes lineal interpolations in a logarithmic space since, at first order, the duration of an evolutionary phase is proportional to the ratio between mass (available fuel) and luminosity (consumption rate). Under the homology assumption, these are related by a power-law dependence.
Again, such an approximation may not be a good one if homology fails.
The potential consequences are clearly illustrated by Fig. \ref{fig:SN}, which shows the supernova (SN) rate obtained using different interpolation schemes. The saw-teeth behavior of the solid line is clearly an artifact of the linear interpolation scheme. The parabolic interpolation
proposed by \cite{Cetal01} yields a much smoother behavior.

\begin{figure}[!ht]
\includegraphics[width=\textwidth]{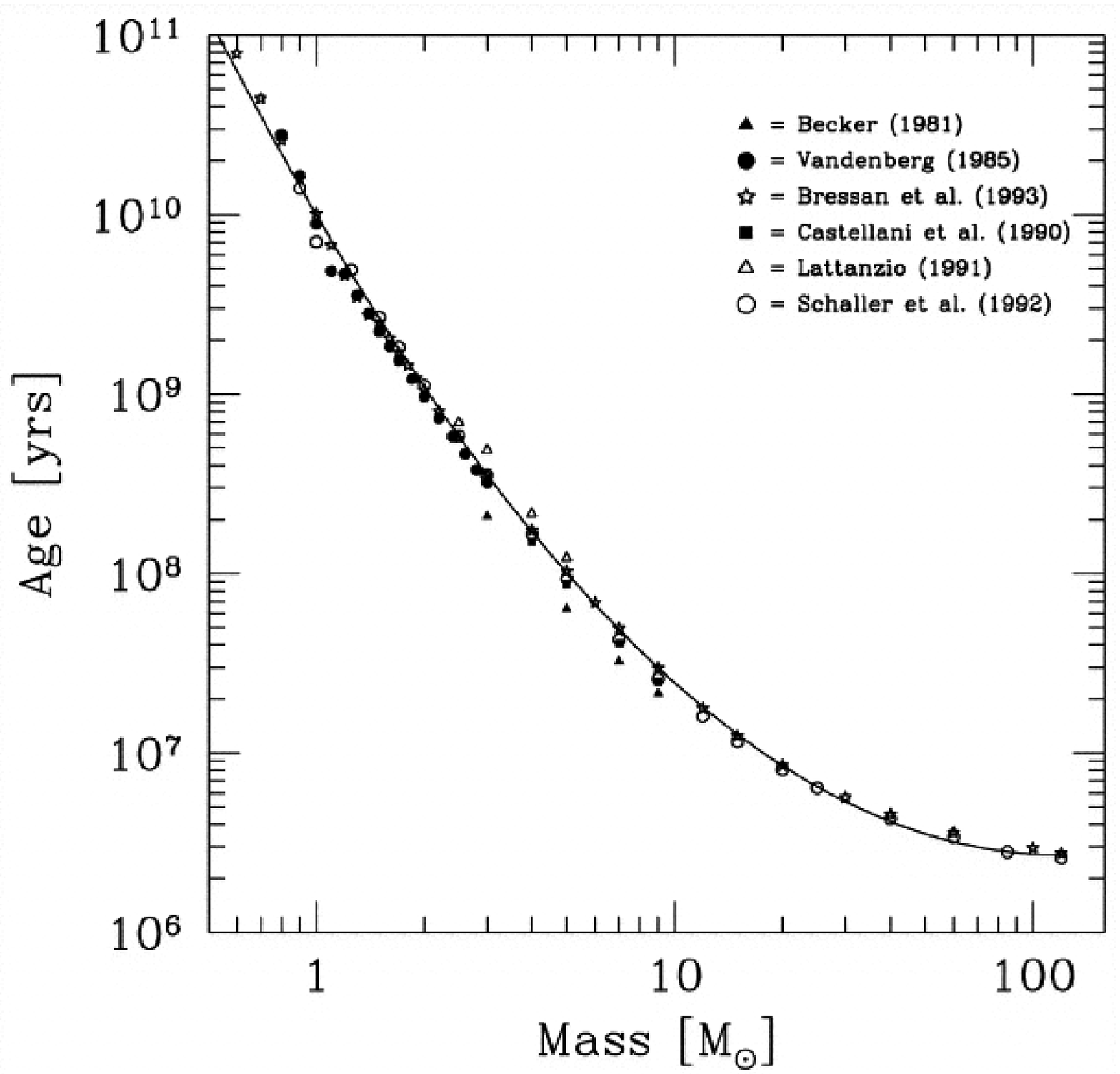}
\caption{Theoretical MS lifetime vs. stellar mass for solar metallicity according to the
evolutionary tracks by \cite{Bec81,Van85,Cas90,Lat91,Schetal92} and \cite{Bre93}.
The solid line is a fit to the data according to Eq. (3). Figure from \cite{Buzz02}.}
 \label{fig:buz}
\end{figure}

The accuracy  of interpolation schemes for stellar evolutionary times has barely been addressed in the literature. In fact, the solution proposed by \cite{Cetal01} is only a second order mathematical approximation, that does not
take the physics of stellar evolution into account. A more accurate 
estimation of the lifetime of different evolutionary phases can be obtained by fitting the
evolutionary times of different evolutionary phases. As an example, we show in Fig \ref{fig:buz} the MS lifetime as a function of the initial mass for different sets of evolutionary tracks. 
The fit of all the points of various sets of tracks  \citep{Buzz02} provides a dependence of the MS lifetime with mass:
 
\begin{equation}
\log t = 0.825 \log^2 \frac{m_\mathrm{TO}}{120} + 6.43.
\label{eq:tms_buzz}
\end{equation}

\noindent This simple formula shows how the interpolation scheme of parabolic interpolations can be greatly improved. 
We refer to \cite{Buzz02} for additional details.

Note that any variation in the time interpolation scheme not only affects the SN rate, but also any other predicted quantity that depend on lifetimes. In particular, the stellar make-up of the synthetic cluster also changes.

\subsubsection{Further inconsistencies in the computed isochrones}

\begin{figure}[!ht]
\includegraphics[width=\textwidth]{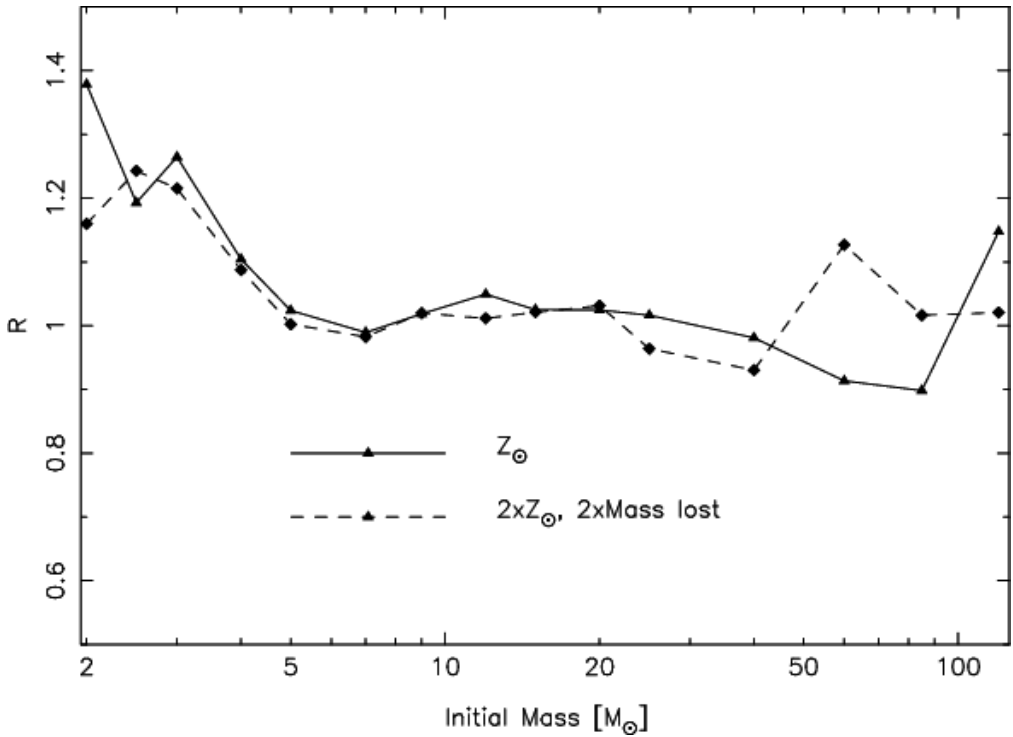}
\caption{Ratio of the integrated mass-loss during the lifetime of the star vs. "reconstructed" mass loss by subtraction of the mass
at the the end of the evolution for the tracks by \cite{Schetal92} (solid line)
at solar metallicity and standard mass-loss rate and the tracks by  \cite{Mey94} at twice solar metallicity and twice mass-loss
rate tracks (dashed line). See \cite{Cetal01} for more details.}
\label{fig:Mdot}
\end{figure}

As we have previously seen, all types of synthesis codes rely on interpolations among tabulated evolutionary tracks independently of the method they use. 
Usually, these interpolations are linear in a logarithmic space for quantities which
are assumed to follow homology relations, such as luminosity or effective temperature.
This interpolation scheme is applied, without any precise physical reason, to other relevant quantities also:
for example, abundance ratios in stellar atmospheres (needed to determine the presence of WR stars in young populations), 
or the mass loss rates (needed to compute the amount of kinetic energy deposed in the interstellar medium).

To illustrate the spurious effects of such {\it ad hoc} interpolation schemes, we show in Fig. \ref{fig:Mdot} 
the ratio $R$ between the mass ejected by a star as obtained from integration of the mass loss rate, and the difference between the initial and final masses as given by the track \cite[see][for more details]{Cetal01}:

\begin{equation}
R = \frac{\sum_{k=1}^{k_e} \dot{m}(t_k) (t_k - t_{k-1})}{M - m_{k_e}},
\end{equation}

\noindent where the index $k_e$ refers to the last tabulated point in the track. 
Consistency would require such a ratio to be equal to 1. However, 
R is found to take values of up to 1.4, i.e. the integrated mass lost by stars is
inconsistent with the 'structural' value by up to 40\%.

\subsection{Uncertainties related to atmosphere libraries}\label{sect:uncertainty_atmospheres} 

As mentioned in \S\ref{sect:atmospheres}, stellar libraries are needed to transform the theoretical space ($T_\mathrm{eff}$, $g$ and $\ell$) into the observational one (colors and/or spectral energy distributions).
Usually, the available model atmospheres form a coarse grid in the
(log $g$, log $T_{\mathrm eff}$) plane, whereas isochrones are
generally continuous in the plane. In order to assign 
a model atmosphere to each isochrone location, 
one can either choose the nearest atmosphere of the grid (closest-model approximation),
or interpolate between nearby atmospheres. Assigning the nearest
atmosphere implies a further binning of data and may originate
jumps in the results. 

\begin{figure}[!ht]
\includegraphics[width=\textwidth]{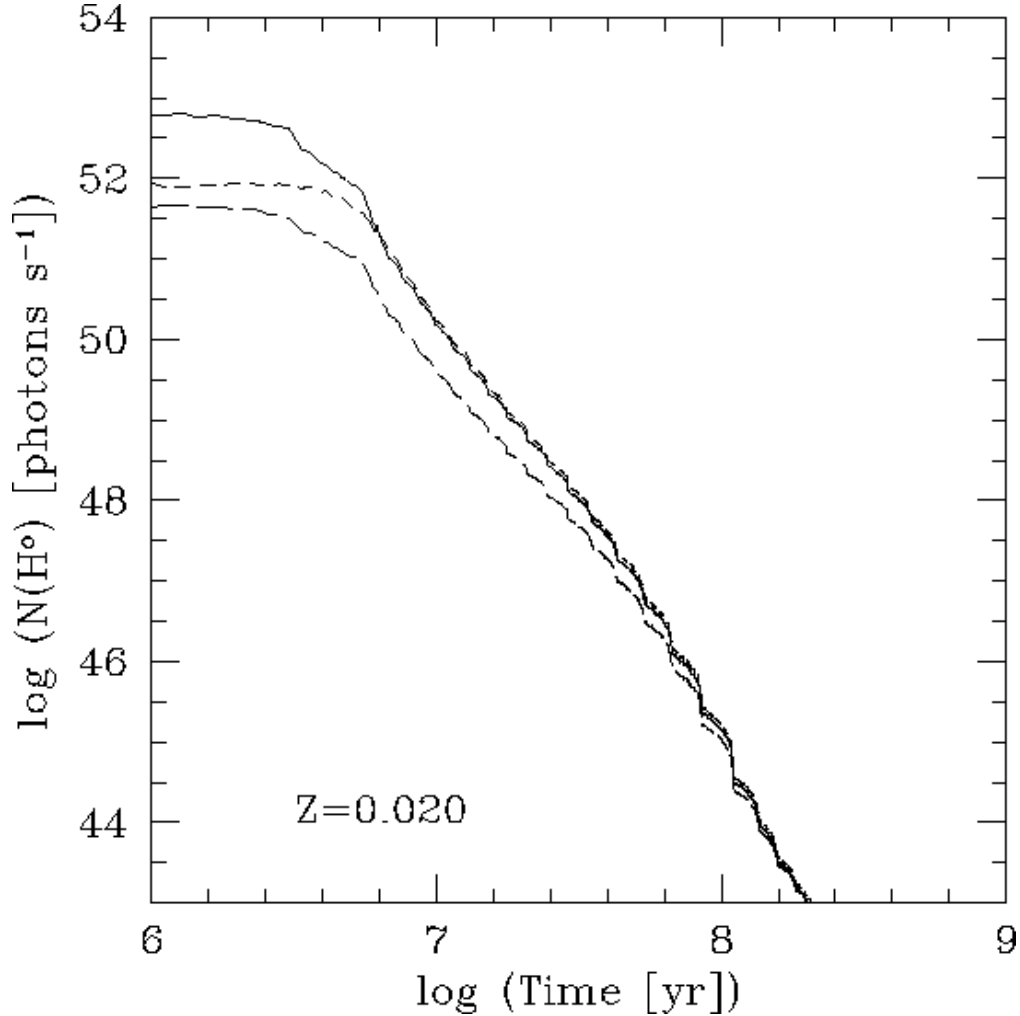}
\caption{Number of ionizing photons as a function of age for solar metallicity and different IMF slopes. Note the downstairs-like behavior  of the quantity at evolved ages due to the transition among atmosphere models. Figure from \cite{SB99}.}
\label{fig:Nlyc}
\end{figure}

It is often assumed that these jumps 
cancel on average when a whole population is considered. Unfortunately, this
is not always the case, as is made clear by Fig. \ref{fig:Nlyc}, which shows the evolution of the number of ionizing photons in a cluster
as a function of age.
The downstairs-like behavior of the plotted  quantity at evolved ages is due to the transition among atmosphere models arising from the use of  the closest-model approximation.

An alternative solution is interpolating among atmosphere models to obtain smoother results.
This solution is not common in synthesis codes, since it increases the computational time. 
However, it is mandatory in the analysis of color-magnitude diagrams (CMD), and, as we will see later, 
it will also be mandatory in synthesis models including rotation. Therefore, 
we mention here some of the unsolved issues related to this kind of interpolations.

If the interpolation scheme is based on the Stefan-Boltzmann law (Eq. \ref{eq:Lbol}) the  interpolations are wavelength independent, and it is simple to perform them linearly in the $\log {\ell}_\lambda - \log T_\mathrm{eff}$ plane. However, for the case of a spectral energy distribution it is more realistic to consider the emission as a black-body spectrum:

\begin{eqnarray}
\log \ell_\lambda(T_\mathrm{eff}) & = &  \log \left(\frac{2\pi h c}{\lambda^5}\right) - \log (e^x -1) = \nonumber \\
& = &  \log \left(\frac{2\pi h c}{\lambda^5}\right) - (x + \log (1- e^{-x}), 
\label{eq:BB}
\end{eqnarray}

\noindent where $x=hc/\lambda k T_\mathrm{eff}$. This means that, at a given wavelength, $\log \ell_\lambda$ is a linear function of the variable $y=\log (e^x -1) = x + \log (1- e^{-x})$. 
Such interpolation scheme clearly differs from the one based on the Stefan-Boltzmann law.
In the intermediate case, i.e. the computations of 
broad or narrow bands, the interpolation scheme should be flexible enough to tend towards both extreme schemes (Stefan-Boltzmann law and black-body law) depending on the width of the band. 
We refer to the appendix in \cite{Jametal04} for a detailed description of this issue.

\subsection{Incompleteness of the input ingredients}

\subsubsection{Incompleteness of evolutionary tracks}

The lack of homogeneous sets of stellar models that cover all the evolutionary phases
is perhaps the most severe problem in population synthesis.
When a synthesis model is to be applied to observations,
the lack of homogeneous stellar models must be sidestepped in some way,
sometimes sacrificing homogeneity, sometimes using inputs based on physics that is known to be incorrect.
Here are some examples: (1) The inclusion of the post-AGB phase in synthesis codes requires using tracks computed by different authors (with possibly different input physics), and performing some {\it ad hoc} assumptions to link the post-AGB to the AGB tracks.
(2) Stellar models computed with enhanced mass loss rates are assumed {\it ad hoc} to mimic the effects of rotation
in massive stellar evolution, although observations point towards mass-loss rates even lower than the standard ones. 

In such cases, assessing the uncertainty in the results is next to impossible,
particularly due to the lack of a physical picture giving a realistic guide of the problem. 
However, these are the only way of producing a complete result when the models are computed for comparison 
with the observations, since they include all the relevant phases of stellar evolution.
On the other hand, when the goal of a synthesis model is to link our (incomplete) knowledge on stellar theory 
to the emission from an ensemble of stars, incomplete input data can be used.
In this case, the results will be incomplete, but they will 
at least predict a lower limit to the expected emission. 

\begin{figure}[!ht]
\plottwo{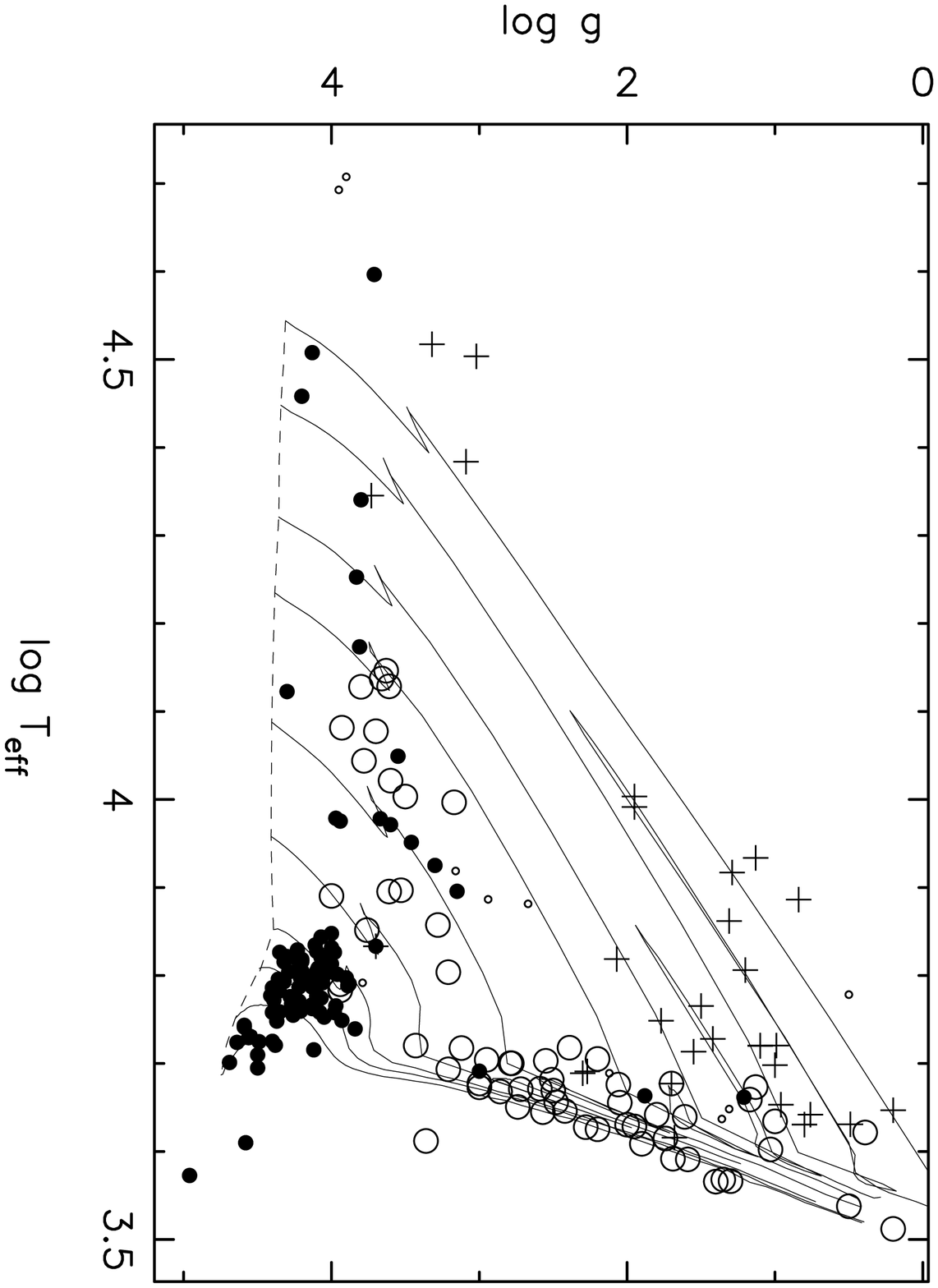}{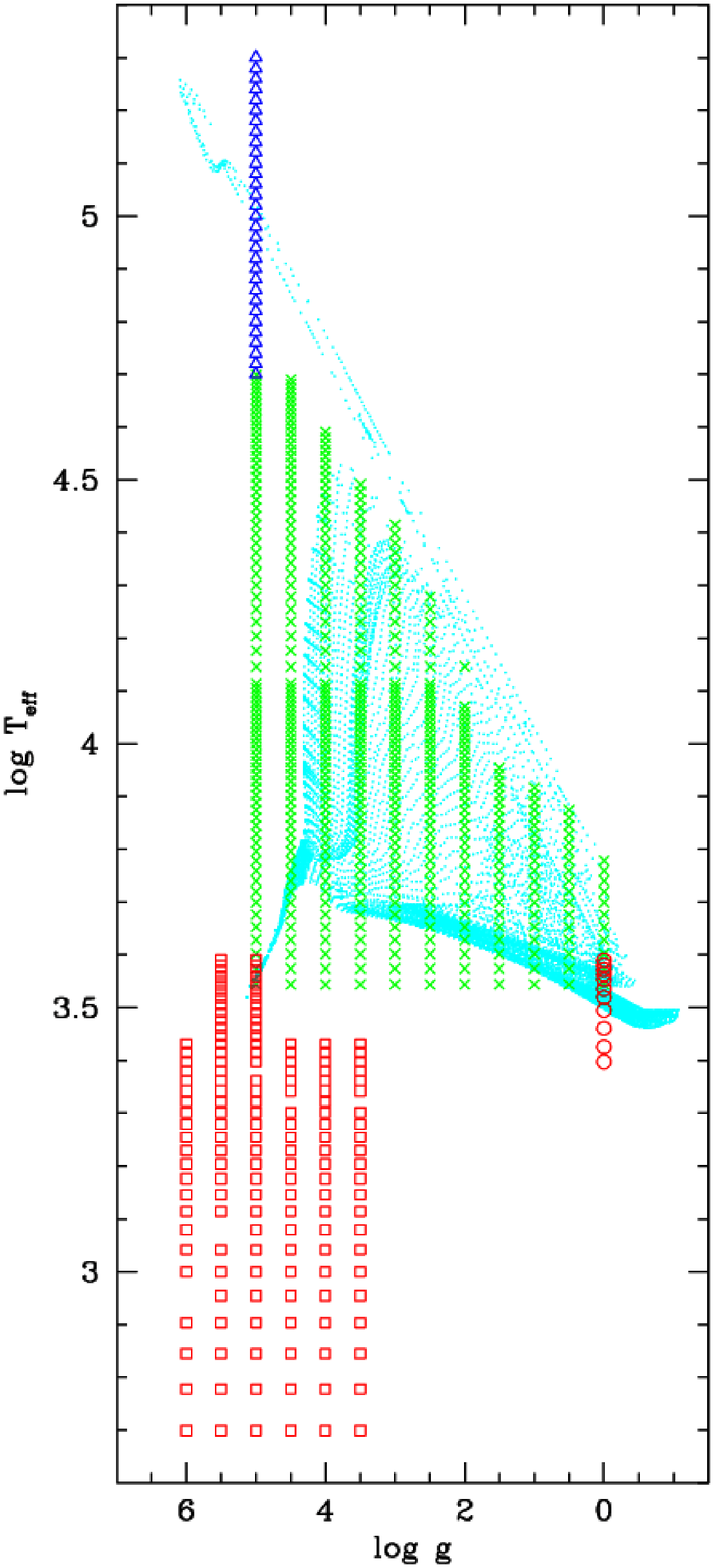}
\caption{{\it Right:} $\log g$ vs. $\log T_\mathrm{eff}$ coverage of the STELIB empirical library \citep{LBetal03} used by \cite{BC03}. The different symbols represent different stellar spectral classes: full circles are dwarf MS stars (class V), open circles are giants (class III), and plus signs are supergiants of classes I and II. Small circles are used for stars with no spectral class determination. {\it Left:}  Stellar library (large symbols) in the   plane $\log g$ vs. $\log T_\mathrm{eff}$, compared to the position of stellar models of solar metallicity (small dots) used by \cite{Gi02}. The spectra are taken from \cite{Casetal97} (crosses), \cite{Fluetal94} (circles) \cite{Alletal00} (squares), and pure blackbody (triangles).  \cite[See][for details]{Gi02}. Figure from \cite{Gi02}.}
\label{fig:atmmod}
\end{figure}

\subsubsection{Incompleteness of atmosphere templates}

The region covered by the isochrones in the $\log g - \log T_\mathrm{eff}$ plane is not the same as the one covered by the atmosphere libraries. Specifically, theoretical libraries do not cover some important regions of this plane, like the one occupied by stars with $T_\mathrm{eff}$ around 10$^4$ K and intermediate $\log g$ values, that corresponds to the A supergiant region,  and the one occupied by cool stars ($T_\mathrm{eff}$ $<$ 3000 K) with negative $\log g$ values, that corresponds to the red supergiant region. To bypass this problem, some codes make use of empirical libraries. 

A severe limitation of empirical libraries is that they do not have a good coverage in metallicity, a fact 
which limits the possibility of applying the models. 
Furthermore, although the red supergiant region is well populated (since these are luminous stars), the coverage in other regions of the $\log g - \log T_\mathrm{eff}$ plane is not sufficient to produce realistic results even at solar metallicity. 

In Fig. \ref{fig:atmmod} the different areas of the $\log g$ vs. $\log T_\mathrm{eff}$ diagram covered by empirical and theoretical libraries can be seen: note how neither of them covers the whole diagram.
Given these limitations, it is impossible for a code to produce a physical self-consistent result.

\subsubsection{Inadequate sampling of the evolutionary phases}

A further problem for the computation of synthesis models is the insufficient sampling of the stellar evolution
along the tracks.
This problem has a different impact depending on the scope of the synthesis model
and the way it is compared to the observational data.
In the case of population synthesis, one wants to reproduce the total luminosity of the cluster
based on the integrated emission, so one needs to take into account carefully the emission
of all the stars.
On the other hand, when studying CMDs one wants to reproduce the emission of the stellar population 
based on its representation in the HR diagram; hence, not all phases are necessary.
This difference has the following consequences:

\begin{itemize}
\item In the case of CMD it is not necessary to cover all the HR diagram, but only the areas under study.
Relevant points in such areas should be included explicitly in the isochrones since
they can be directly compared with individual stars. This is the case of the TO, the tip of the RGB or the tip of the AGB. These points are usually included in the evolutionary tracks and the isochrones.

\item In the case of population synthesis the objective is to obtain the integrated light from the ensemble. So suitable isochrones should be defined by points that are a correct representation of all the evolutionary phases, which are not necessarily the extreme points. For example, if a stellar class is defined around the mass point of maximum luminosity, the corresponding contribution is overestimated.

\item Not all the stellar models computed to follow the stellar evolution are eventually published (Georges Meynet and  Daniel Schaerer, private communication; Antonio Claret, private communication). The reason is that, in regions where real stars are barely observed, the advantage of including these points is not considered worth
the extra-memory needed. Hence, evolutionary tracks are reduced to a minimal set of equivalent evolutionary points plus a few additional points that are relevant for stellar evolution theory.
However, population synthesis would require the detailed knowledge of the evolutionary path in the HR diagram.
\end{itemize}

Summing up, evolutionary tracks are optimized for the comparison with CMD (which provides a  first reliable check of the assumed physics), but not to perform evolutionary synthesis studies.
These difference is clear when extreme phases of stellar evolution are considered. For example, the tip of the RGB is a useful point for representation in CMD, but it is not a suitable point for population synthesis, since it represents a transient high luminous phase in stellar evolution that pollutes 
the emission of the ensemble. As a further example, SN explosions are not included in synthesis models because they dominate the emission with their high luminosity (i.e. the resulting model would have a spectral energy distribution that corresponds to a single component, the SN explosion, whatever the underlying population is).

The previous conclusion can be restated in a different way: since synthesis models are designed to obtain the emission of an ensemble of stars, and not individual stars, their use 
should be limited to situations where an ensemble really exists. 
A necessary condition to fulfil this requirement has been formulated by \cite{CL04} by defining the {\it Lowest Luminosity Limit} that an observed cluster must have 
in order to be modelizable with synthesis codes:

\begin{center}
{\it The total luminosity of the observed cluster must be larger than the individual contribution of any of the stars included in the model.}
\end{center}

Quite obviously, the inclusion of SN events in the model increases the corresponding Lowest Luminosity Limit.

\section{Uncertainties related to the numerical methods}\label{sect:numerical}

The final source of uncertainty in synthesis models lies in the numerical methods used. 
Three types of uncertainties arise at this level: the presence of bugs in the program, the accuracy of the numerical methods, and the consistency of the algorithm used.

\subsection{Bugs}

\begin{figure}[!ht]
\includegraphics[width=\textwidth]{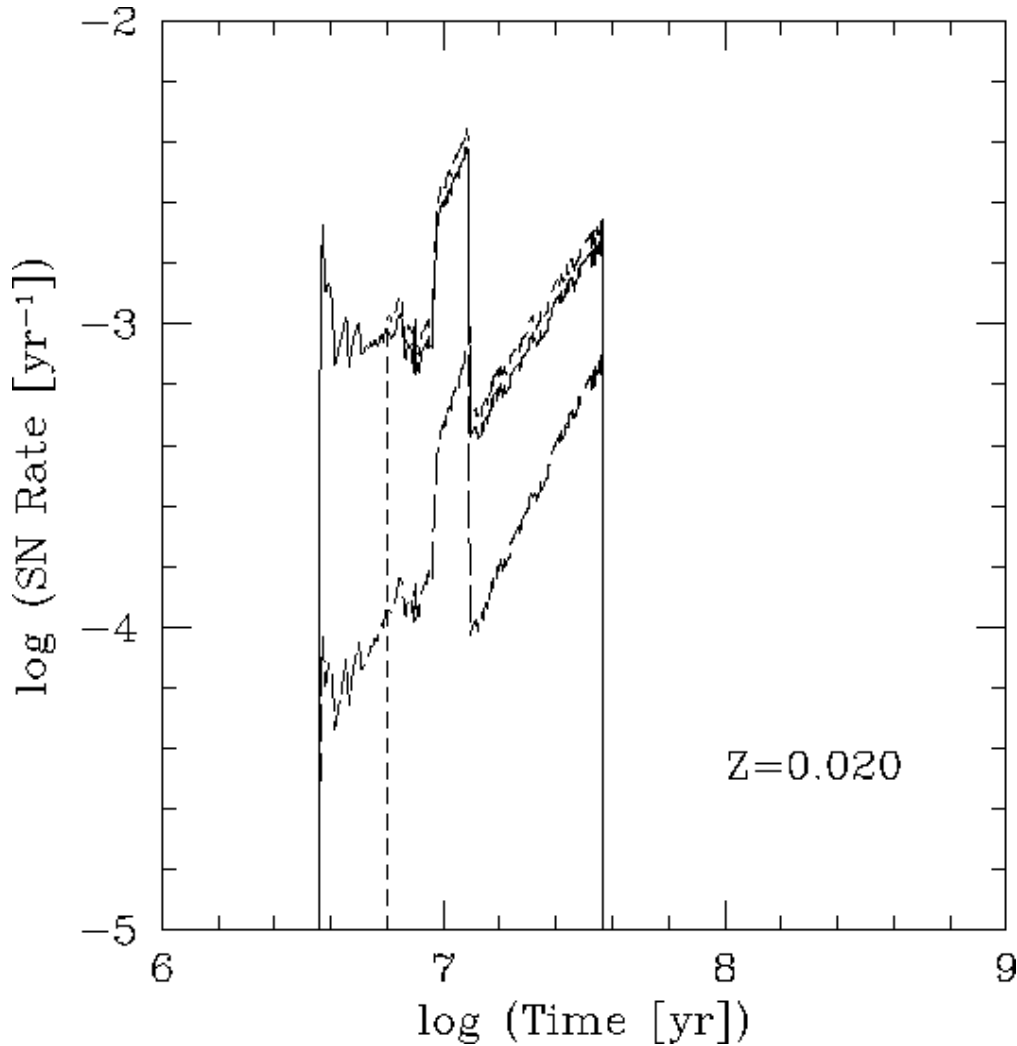}
\caption{SN rate vs. time obtained by the original version of Starburst99 code. Star formation law: instantaneous; solid line, $\alpha = 2.35$, M$_\mathrm{up}$ = 100 M$_\odot$; long-dashed line,
$\alpha  = 3.30$, $M_\mathrm{up}$ = 100 M$_\odot$; short-dashed line, $\alpha  = 2.35$, M$_\mathrm{up}$ = 30 M$_\odot$;  Z = 0.020.
This figure can be compared with Fig. \ref{fig:SN}. The bug was fixed in Starburst99 code following the prescriptions shown in \cite{Cetal01}.}
\label{fig:SNSB99}
\end{figure}

As observed by Ferland (http://www.nublado.org), the presence of bugs is inevitable in any large code.
Sometimes they are documented and solved,
 and corrected in following versions. As an example, we show here the SN rate obtained by the original version of Starburst99 (c.f. Fig \ref{fig:SNSB99}). The figure can be compared to Fig. \ref{fig:SN}, which is a correct representation of the SN rate. Note in particular that the wrong SN rate increases with time, whereas the corrected one
tends to decrease.

The bug was fixed in Starburst99 following the prescriptions by \cite{Cetal01}.
The source of this bug was an incorrect determination of the $m_i^\mathrm{low}$ and $m_{i}^{\mathrm{up}}$ limits in Eq. \ref{eq:wi}. Although the bug showed up in the SN rate, it also affected the $w_i$ determinations of all the other evolutionary phases. The effect of the bug was particularly evident in the case of the SN rate because,
being this a derivative, it amplifies any discontinuity in the primitive function.

\subsection{Accuracy}

\begin{figure}[!ht]
\includegraphics[width=\textwidth]{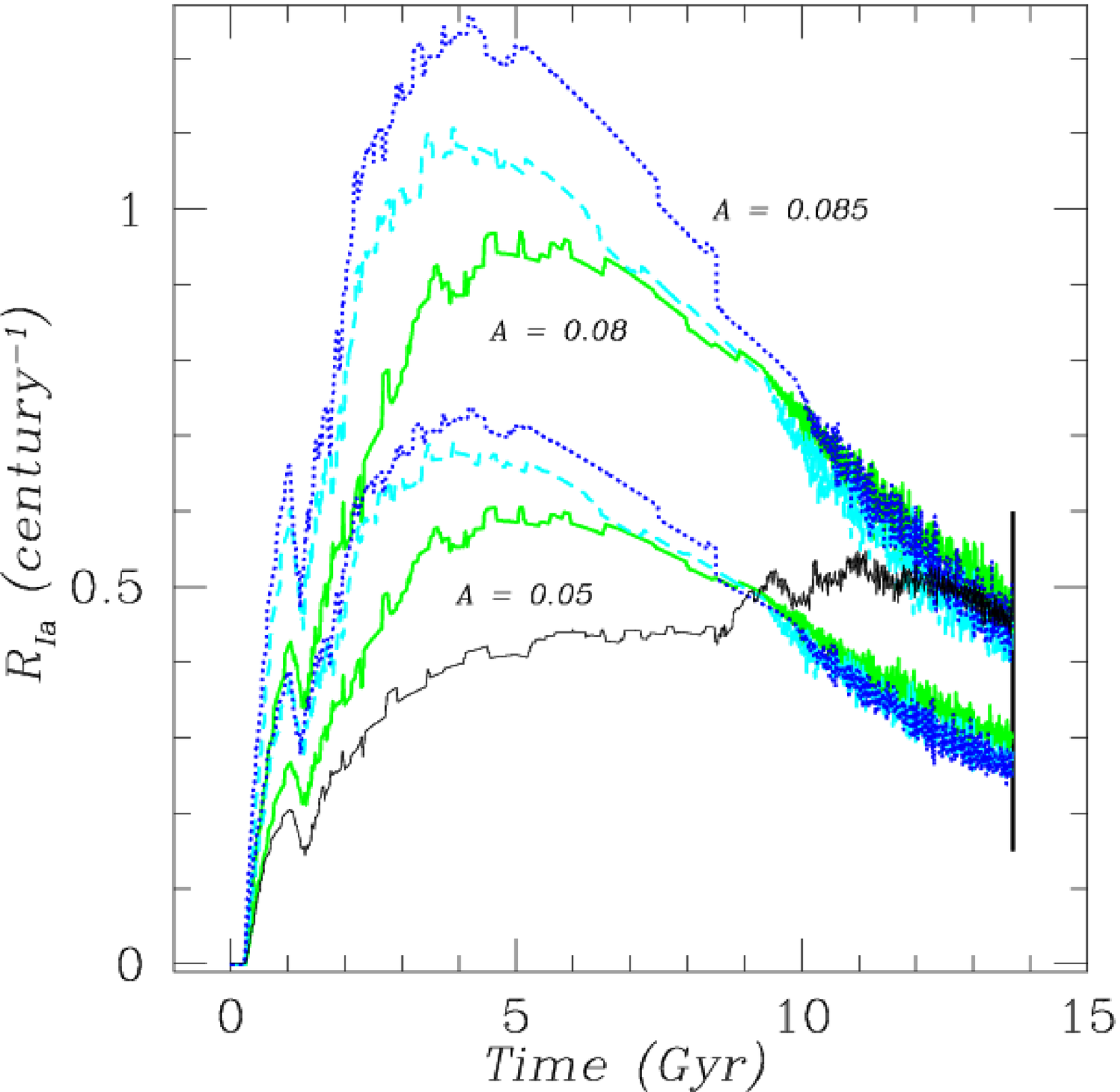}
\caption{Type Ia SN rates obtained with different assumptions for the
stellar lifetimes: \cite{MM89} (thin solid line); \cite{Tin80} (thick dotted line); \cite{Schetal92} (thick solid line); \cite{Kod97} (thick dashed line). The effect of changing the
fraction of mass entering the formation of type Ia SN progenitors, $A$,
is also shown.
The type Ia SN rate observed in the Galaxy at the
present time is also shown \citep{Cappetal97}. See \cite{Rometal05} for more details.
Figure from \cite{Rometal05}.}
\label{fig:SNI}
\end{figure}

The second source of uncertainty is the accuracy of the numerical methods used. As an example we plot in Fig. \ref{fig:SNI} the SN rate of Type I SN computed by \cite{Rometal05}. In this case, the plot shows a correct overall behavior. However, at age larger than 10 Ga,  an oscillation of the results around a mean value is observed. This behavior is caused by insufficient numerical precision.  This oscillation had only a minor effect on the original paper, which deals with chemical evolution, a field in which the relevant quantities come from cumulative contributions; but if this model were applied to stellar population synthesis, the nominal tabulated result could not be used.

\subsection{Algorithms}

\begin{figure}[!ht]
\includegraphics[width=\textwidth]{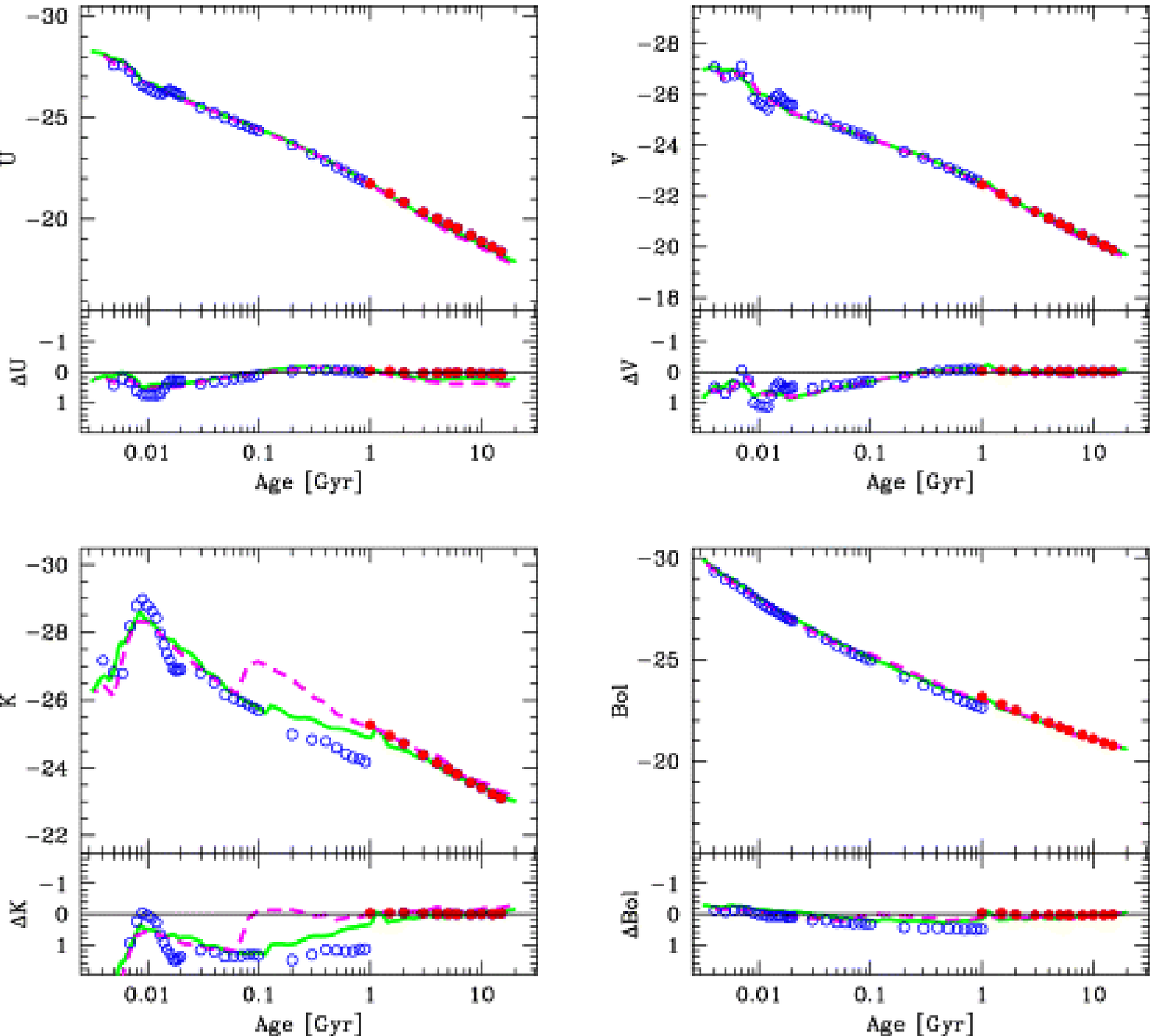}
\caption{Upper plots of each panel: luminosity evolution of the SSP models by \cite{Buzz05}  (solid dots), for solar metallicity and Salpeter IMF, compared with other theoretical outputs according to \cite{SB99} (open dots) \cite{Breetal94} (dashed line) and \cite{BC03}  (solid line). The total mass is scaled to M$_\mathrm{SSP}$= 10$^{11}$ M$_\odot$ throughout, with stars in the range 0.1$-$120 M$_\odot$. The \cite{SB99} model has been slightly increased in luminosity (by about 0.06 mag in bolometric luminosity, at 1 Ga) to account for the luminosity contribution of low-MS stars (M$* \leq 1.0$ M$_\odot$). Lower plots of each panel: model residuals with respect to SSP fitting functions. Figure from \cite{Buzz05}.} 
\label{fig:buzz05}
\end{figure}

FCT and isochrone synthesis codes have been shown to produce systematically discrepant results
\citep{CB91}. As explained in \S\ref{sect:types}, the two types of codes use different algorithms
to compute the contributions of the different stellar classes. However, it must be emphasized that
 both types of codes are based on the same underlying physics. 
Furthermore, the FCT can be taken into account in the isochrone computations, as shown by \cite{Breetal94} and, especially, \cite{MG01}.
Hence, both methods should produce exactly the same results, provided they have, of course, the same inputs. Unfortunately this is not the general case, a fact that suggests that the differences among synthesis codes computations could be due to differences in the interpolations scheme.
Note that the current market of synthesis codes is dominated by isochrone synthesis codes that compare
their results with similar models. Such comparisons do not shed any light on this problem,
but only provide a consistency test for the mathematical methods used by isochrone synthesis. Only comparisons among
different methods will increase the reliability of evolutionary synthesis:
{\it the stellar population synthesis method will not be a reliable tool until the isochrone and the FCT methods yield consistent results.} 
Although this may seem a pessimistic point of view, especially to those who make use of synthesis codes to infer the physical properties of observed clusters,  the situation is not so dramatic:
in spite of persisting differences in the results, there are also regions of the electromagnetic spectrum where both methods provide similar results \cite[see][ and Fig. \ref{fig:buzz05} for a comparison]{Buzz05}. Hence, such wavelength regions can be confidently compared to observational data.

\subsection{Documentation, documentation, documentation}

In this section we have shown that, besides the possible uncertainties in the input ingredients, there are
also uncertainties associated to the methods used. The evaluation of such uncertainties is quite difficult since they depend on the code itself.
The only feasible way to evaluate how reliable the model results are is to write an exhaustive documentation of the corresponding code, both at the level of a user's guide and
 at a technical level. A good example to follow is the documentation
of the photoionization code CLOUDY ({\tt http://www.nublado.org}).

\section{Rotation and variability}\label{sect:future}

In the previous sections we have addressed the uncertainties in synthesis models 
in the  simplest   and most idealized case which is
the assumption of the existence of a {\it single} isochrone that defines the stellar mixture at a given age.  But in a stellar cluster it is possible to have stars with distributed features, such as stars that rotate with different rotational velocities, variable stars or binaries, such that their overall emission is also
distributed.
In terms of modeling, this means that a well-defined spectral energy distribution (the main output of synthesis models) is not enough to characterize the cluster, and that a distribution range must also be provided.

\begin{figure}[!ht]
\includegraphics[width=\textwidth]{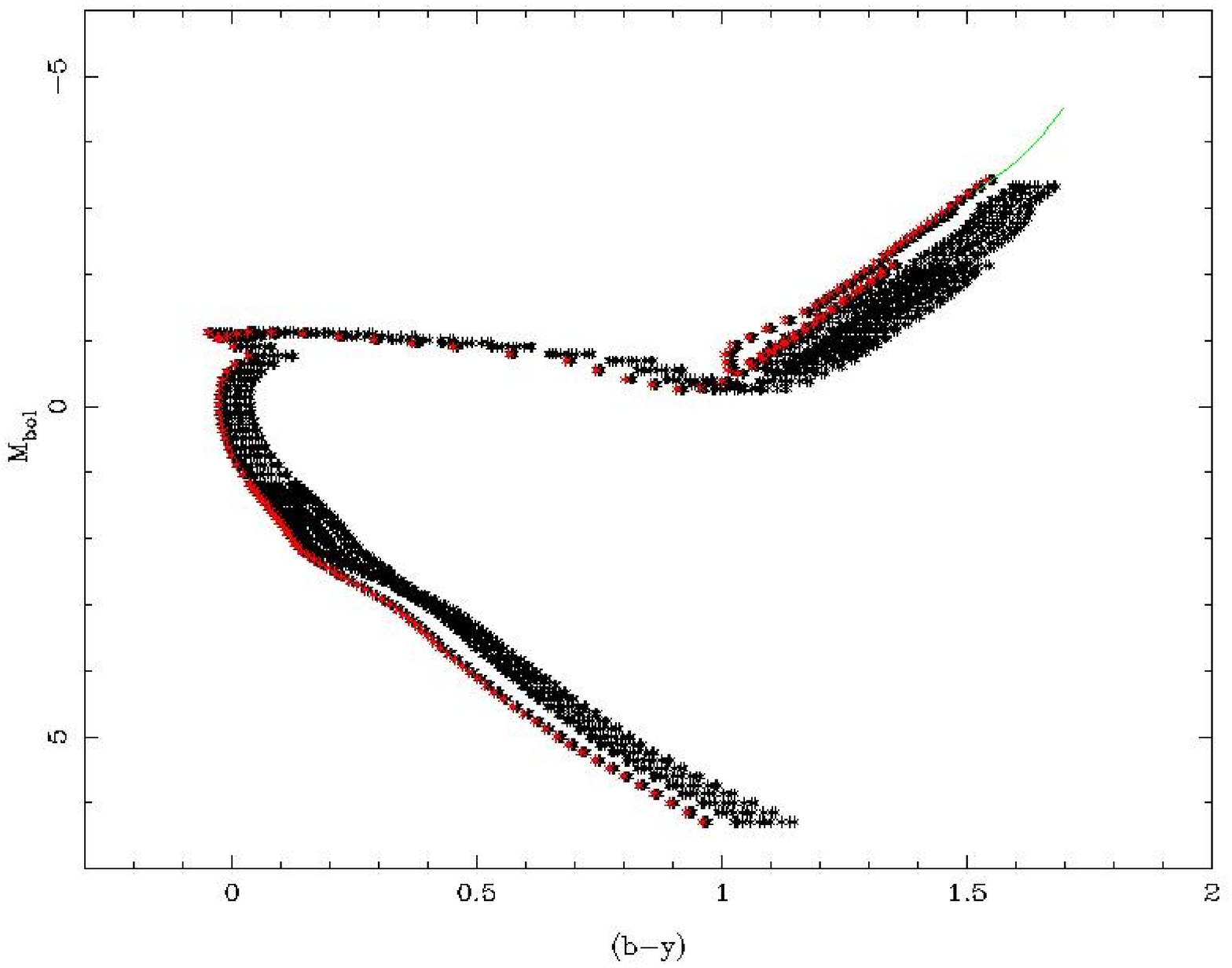}
\caption{HR diagram for isochrones at the same age  with different rotational velocities and inclinations angles. Courtesy of Antonio Claret}
\label{fig:bin}
\end{figure}

The case of rotation is illustrated in Fig. \ref{fig:bin} where a set of isochrones at the same age is plotted. Rotation produces a loss of symmetry in the star, so that its emission depends on the inclination angle of the star with respect to the observer. Points in Fig. \ref{fig:bin} differ from each other in the initial mass, the rotational velocity and the inclination angle \cite[see][for more details]{Cla00,Cla03,CPH05}.

The inclusion of rotating stars in synthesis models will imply a global revision of synthesis models that includes:

\begin{itemize}
\item Determining whether homology relations exist for rotating stars, in order to interpolate tracks and compute isochrones.
\item Studying  the distribution of {\it initial} rotational velocities and including it in the stellar birth rate.
This also means making additional assumptions about whether the rotational velocity distribution and the IMF are separated distribution.
\item Including an inclination angle distribution in the stellar birth rate. Such a distribution should be a random (flat) distribution.
\item Abandoning the closest atmosphere approach in favor of an atmosphere interpolation scheme, and studying the variation of line profiles with rotation (for high-resolution atmosphere templates): all the advantages of including rotation in the isochrones will be lost if, at the end, the combinations of spectral energy distributions do not take into account the richness of available situations and the isochrone points are described by a discrete set of atmosphere templates.
\item Obtaining not only average results but also the uncertainty associated to the mixture of distribution angles in the cluster.
\end{itemize}

\section{What is really computed?}\label{sect:discussion}

Although all evolutionary synthesis codes pursue essentially the same goal, they may differ substantially from each other in several aspects. Here, we will discuss two major distinctions within the field, the first related to the specific procedure followed in populating the IMF and the second to the interpretation of results. 

\subsection{Modeling philosophies} 

Evolutionary synthesis models can be either probabilistic or deterministic in two distinct aspects: a) the IMF sampling, and b) the interpretation of results.

\subsubsection{The IMF sampling}

\begin{figure}[!ht]
\includegraphics[width=\textwidth]{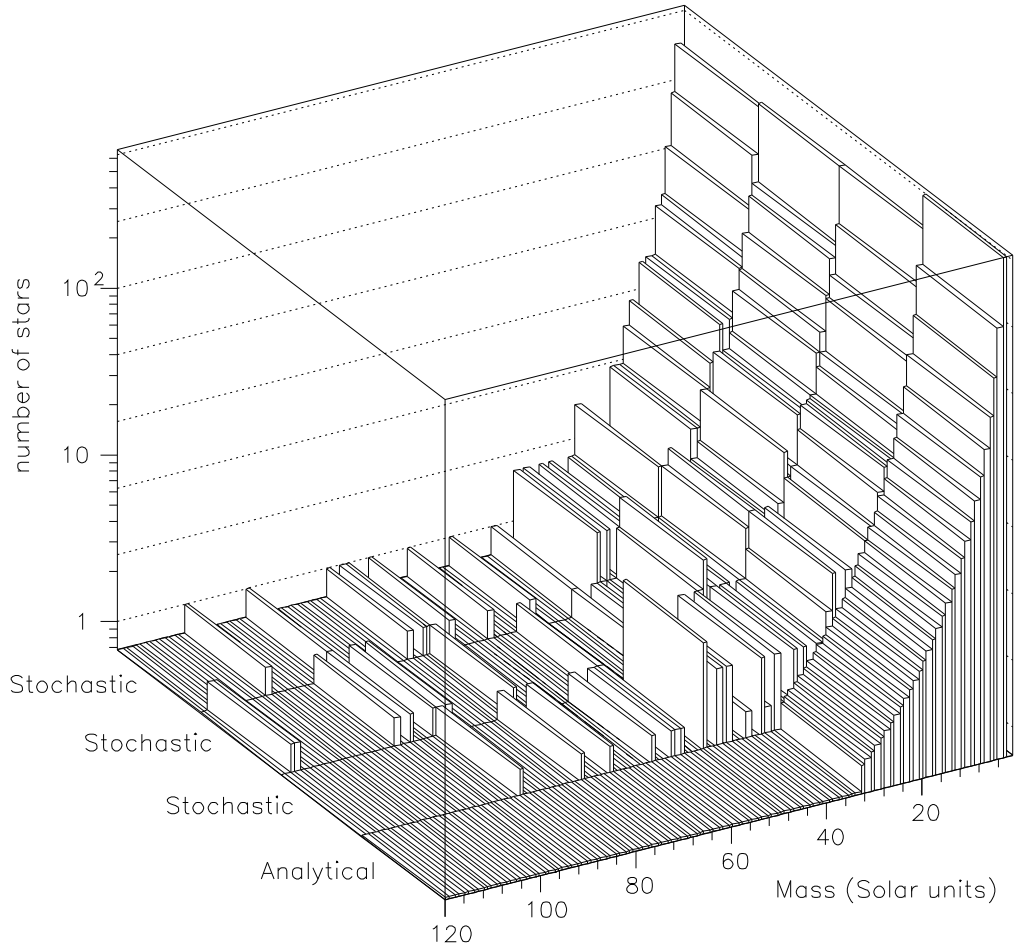}
\caption{IMFs computed with a slope $\alpha=2.35$ for 10$^3$ stars in the mass range 2-120 M$_\odot$. It is compared here three stochastical Monte Carlo simulations with an analytical function. Figure taken form \cite{CMH94}.}
\label{fig:CMH94}
\end{figure}

In a synthesis code, the IMF can be used either as an exact or as a probabilistic description of the mass distribution. In the first case, the relative frequencies of newborn stars of different masses are univocally fixed by the IMF; in the second, the mass of each newborn star is assigned through a random selection process where the IMF is the weighting function. In the first case, if the IMF is continuous, a smooth (although binned) distribution of stars results; in the second, the resulting distribution will be noisy, and the smaller the number of stars in the stellar population, the noisier the distribution. Two runs of the code with the same input parameters will yield two identical models in the first case, but not in the second (Fig. \ref{fig:CMH94}). The first approach is adopted by the so-called standard codes while the second is adopted by Monte Carlo codes.         

Standard codes are sometimes called analytical, but this adjective is misleading, because: a) in either case the underlying IMF may be an analytical function, and b) although the IMF is considered analytical, a discretization step (binning) is required by both methods. Hence, both methods are analytical in the same measure, and we will use here the less ambiguous expression {\it standard}.

\subsubsection{The interpretation of results}

As mentioned before, the main final product of a synthesis model is the expected value of the total luminosity. What does `expected' mean? Let's see here which are the possible answers. 

\begin{figure}[!ht]
\includegraphics[width=\textwidth]{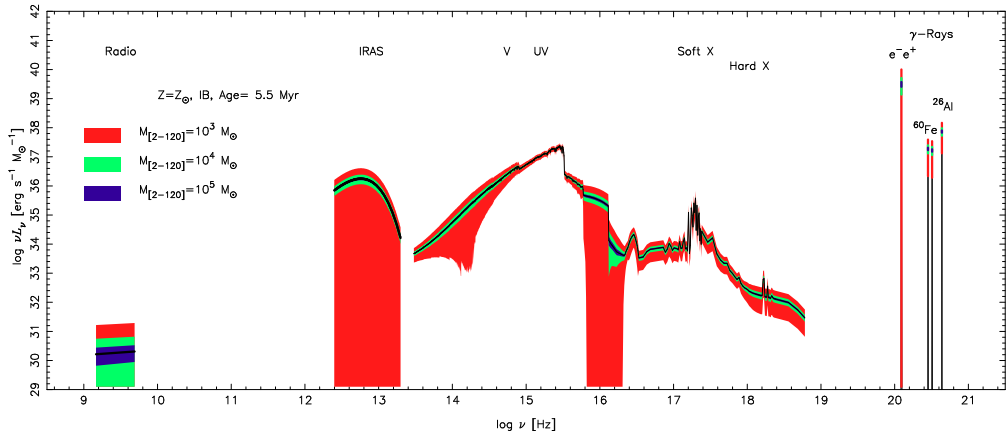}
\caption{Comparison of predicted luminosity of a standard model at different wavelengths with the 90\% confidence interval predicted by Monte Carlo models of 2. 10$^5$ M$\odot$ (blue), 3. 10$^4$ M$_\odot$ (green) and 3. 10$^3$ M$\_odot$ (red). Models computed with sed@ code \citep{Cgam00,CVGLMH02,CMHK02}. Figure from \cite{Cer03} }
\label{fig:multi}
\end{figure}

The interpretation of the output of Monte Carlo codes is straightforward: the luminosity computed by the code is the particular luminosity of a particular realization of all the possible clusters characterized by a given set of input parameters. Running a set of Monte Carlo models with the same set of input parameters and data will produce a distribution of output luminosities. This distribution is a sampling distribution of the underlying Integrated Luminosity Distribution Function. The more numerous the stars in each simulation, the narrower the distribution (in relative terms). An example of this trend can be seen in Fig. \ref{fig:multi}. The solid line in the middle of the colored areas is the prediction of a standard model at different wavelengths; the colored bands are the areas spanned by sets of Monte Carlo models with the indicated number of stars. A striking feature of this plot is that the uncertainty area depends critically on the wavelength considered.

\begin{figure}[!ht]
\includegraphics[width=\textwidth]{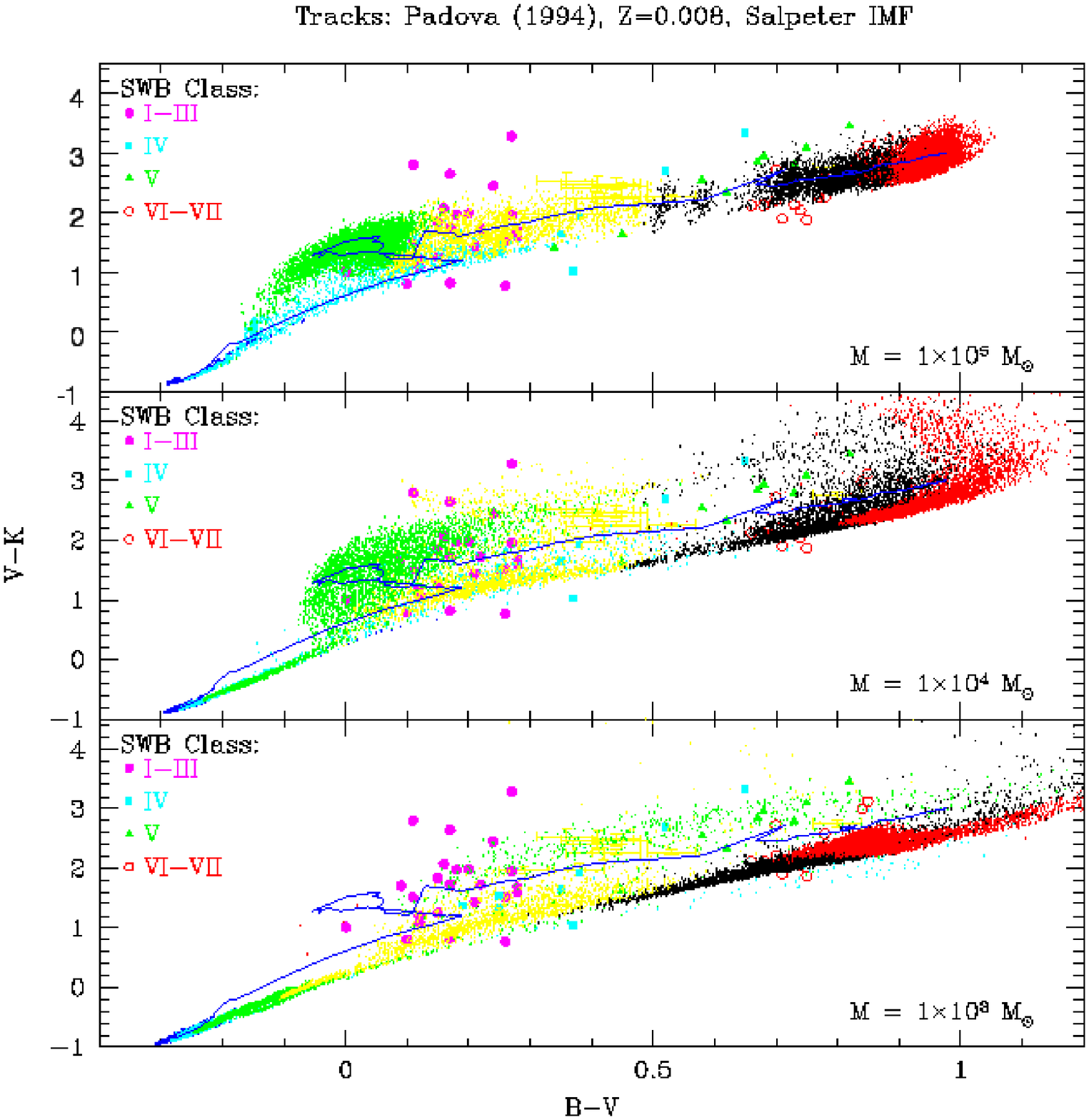}
\caption{Comparison of Monte Carlo simulations (small dots), analytical simulations (lines) and observations of LMC clusters (medium dots). Models has been performed by the code presented by \cite{BC03} with \cite{Fetal94} evolutionary models and \cite{LCB98} atmosphere models. Figure from \cite{Brutuc}.}
\label{fig:BruTuc}
\end{figure}

Another example is found in Fig. \ref{fig:BruTuc}, where  the U$-$B and B$-$V colors
resulting from different simulations in which stochastic fluctuations in
the number of stars that populate different evolutionary stages is plotted with small dots. Lines corresponds to analytical simulations and medium dots to  LMC clusters. The models have been computed with the code by \cite{BC03}. 
From this figure it can be seen that Monte Carlo simulations for small clusters lie in a region that is not covered by the analytical simulations. This bias in colors is a natural effect when the 
integrated light coming from an observation with a low number of stars  is analyzed \cite[see ][for more details]{CVG03}.

In the case of standard models, however, different meanings can be attached to the output luminosity. Each time we run the code with a given set of input parameters we obtain the same result, so in this sense standard models are deterministic. Indeed, the traditional interpretation of standard synthesis models is wholly deterministic: the output is interpreted as if it were the luminosity of an (ideal) cluster with the same characteristic of the model. Models of this type, with their associated interpretation, are called deterministic synthesis models. 

Such interpretation is based on the assumption that the IMF is well sampled, which in turn implies that the IMF is seen as an exact description of the stellar population. This assumption is questionable, since whatever the mechanism of star formation is, discrete entities cannot map continuously a curve. Even if we divide the mass range in bins, a continuous distribution would eventually imply fractional numbers of stars in certain bins, which is unphysical. On the other hand, if we allow for a certain degree of stochasticity in the process of star formation (which, on physical grounds, seems plausible), these contradictions are overcome. But allowing for stochasticity in the IMF implies that the overall cluster properties will be stochastically distributed. This view is the one held by statistical synthesis codes.

In sum, deterministic synthesis codes interpret the (deterministically obtained) output luminosity as the total luminosity of the cluster, while statistical synthesis codes interpret it as the mean value of a distribution of possible luminosities. The distinction between the concept of total luminosity and the concept of mean value of a luminosity distribution is a fundamental one: as an example, the mean of a distribution is not necessarily a value that the variable can take. Furthermore, knowledge of the mean value is not enough to estimate even roughly the variable's value, if the shape of the distribution is not known: that is, the mean value is neither an actual value of the total luminosity of a cluster, nor necessarily a good guess of it. 

Statistical and deterministic models are not actually different classes of models, but rather different interpretations of the same class of models, the standard ones. Some standard codes do not explore this interpretation and thus produce purely deterministic models, while others have built-in statistical tools that add to the main result, the luminosity, estimates of the other statistical parameters of the distribution. 

\subsection{Statistical approximations to the problem}

The issue of sampling and the interpretation of the results of synthesis models has been addressed in the literature from different points of view. They include:

\begin{itemize}
\item {\it Monte Carlo simulations} and their comparison to standards models: these works make a description of the problem, but they do not provide any general solution other than performing tailored
Monte Carlo simulations for the cluster under consideration. Examples are: \cite{BB77,Chietal88,SF97,CLC00,LM00,Brutuc,GiTuc,CRBC03}.

\item {\it Theoretical approaches based on the IMF sampling and Surface Brightness Fluctuations studies}: These works try to quantify the impact of sampling effects on the interpretation of the results. Although relevant as a first step towards a characterization of the underlying luminosity functions, they are of limited power when it comes to obtaining the variance of the corresponding distribution. 
In this specific subject there have been two different approaches:

\begin{enumerate}
\item The variance of the luminosity function has been computed in terms of an {\it effective number of stars} $\cal N$, defined as the ratio between the square of the mean and the variance of the luminosity function \citep{Buzz89}:

\begin{equation}
\frac{1}{{\cal N}} = \frac{\sigma^2}{\mu^2} = \frac{\sum w_i l_i^2}{(\sum w_i l_i)^2}.
\label{eq:neff}
\end{equation}

This expression is used in the works by \cite{Buzz89,CVGLMH02,CVG03,GLB04,CL04}, and their use is focused on the determination of the uncertainty of synthesis models.

\item The variance is presented in terms of {\it Surface Brightness Fluctuations}, $\bar{L}$, defined as the ratio between the variance and the mean of the luminosity function \citep{TS88}:

\begin{equation}
\bar{l} = \frac{\sigma^2}{\mu^2} = \frac{\sum w_i l_i^2}{\sum w_i l_i}.
\label{eq:SBF}
\end{equation}

SBF are used to obtain extragalactic distances, and they have also been proposed as a means to break the age-metallicity degeneracy in old stellar populations. They are used, e.g., by \cite{W94,Buzz93,Liu00,Mei01,Blaetal01,GLB04}; see also the contributions by Rosa A. Gonz\'alez and Maurizio Salaris in these proceedings.
     
\end{enumerate}

As \cite{Buzz93} shows, both quantities are different ways to express the same quantity, and they express the statistical entropy of synthesis models (see also this proceedings or astro-ph/0509602). Given that both  $\cal N$ and $\bar{l}$ are estimates of the mean and the variance of the luminosity function, these quantities provides a powerful tool for the comparison of synthesis models. The differences in the $\bar{l}$ value found by different authors \cite[see][for a comparison]{Liu00,Mei01} suggest either a difference among the luminosity functions implicitly computed by synthesis models or problems in the numerical computations. 

\item {\it Theoretical approaches based on luminosity functions.} In other works, the luminosity function or its higher-order moments are studied.
The approach is the only one that can provide confidence intervals of results. Examples are    \cite{GGS04,CLCL05}.

\end{itemize}
    
Although the statistical modeling of stellar populations is not extensively used by the community yet, 
it is expected that this situation will change in the future, since it is the only interpretation that allows to explain systems composed by any number of star, and provide confidence intervals in the application of the models to real observations. The complete statistical formulation will be described in next section.

\subsection{Complete statistical formulation}

In this section, we will explain the basics of the method of statistical modeling. 
For a complete descroption of the method, see~\cite{CLCL05}.
The general problem is the computation of the luminosity of an ensemble of stars. A {\it luminosity distribution function} (LDF: $\varphi_\mathrm{L}(\ell)$) can be assumed, which describes the expected distribution of luminosity 
values in a generic ensemble. The integral of the LDF is normalized to 1:

\begin{equation}
\int_{0}^{\infty} \varphi_\mathrm{L}(\ell)d\ell = 1.
\label{eq:normLDFtrad}
\end{equation}
     
\noindent The integrated luminosity of an ensemble
is traditionally obtained by means of the expression:

\begin{equation}
L_{\mathrm{tot}} = N_{\mathrm{tot}} \int_{0}^{\infty} \ell \, \varphi_\mathrm{L}(\ell)d\ell.
\label{eq:Ltot_int}
\end{equation}

The bottom line of the statistical formulation is that the traditional approach shown in Eq. \ref{eq:Ltot_int}
is conceptually wrong and operationally sterile.
The crucial point here is
the definition and interpretation of the LDF:
$\varphi_\mathrm{L}(\ell)$ is a  {\it probability density function} 
(PDF) from which the luminosities of individual sources are drawn; 
accordingly, the total luminosity of a cluster 
cannot be deterministically computed, but its {\it mean} value $M_1'$
can be obtained as:

\begin{equation}
M_1'  \equiv \langle L_{\mathrm{tot}}\rangle = N_{\mathrm{tot}} \int_{0}^{\infty} \ell \, \varphi_\mathrm{L}(\ell)d\ell.
\label{eq:Ltot_mean}
\end{equation}

The integral on the right-hand side of this equation is the mean value of the LDF, $\mu_1'$, so that:

\begin{equation}
M_1' = N_{\mathrm{tot}} \, \mu_1'.
\label{eq:Mu_Ntotmu}
\end{equation}

In terms of the IMF and the isochrone, the LDF can be expressed as follows:

\begin{equation}
\varphi_{\mathrm{L}}(\ell;t) = \varphi_\mathrm{M}(m) \times \biggl( \frac{d\ell(m;t)}{dm} \biggr)^{-1},
\label{eq:ldf-imf}
\end{equation}

\noindent where the time dependence of $\varphi_{\mathrm{L}}$ has been written explicitly.
If we change the integration variable in Eq. ~\ref{eq:Ltot_mean} from $\ell$ to $m$, 
the mean value of the LDF can be rewritten 
in terms of the isochrone and the IMF\footnote{The isochrone is not monotonic,
so that the integral limits of Eq.~\ref{eq:Ltot_mean} do not correspond to those of Eq. \ref{eq:muIMF}.}:

\begin{eqnarray}
\mu_1'(t) &=& \int_{m^\mathrm{low}}^{m^{\mathrm{up}}} \ell(m;t) \, \varphi_\mathrm{M}(m) \, \biggr(\frac{d\ell(m;t)}{dm}\biggl)^{-1} \,  \frac{d\ell(m;t)}{dm}\, dm  \nonumber = \\ 
               &= &  \int_{m^\mathrm{low}}^{m^{\mathrm{up}}} \ell(m;t) \, \varphi_\mathrm{M}(m)\, dm \simeq  \nonumber \\
               &\simeq & \sum_i  w_i \,\ell_i(t),
               \label{eq:muIMF}
\end{eqnarray}

\noindent where ${m^\mathrm{low}}$ and ${m^{\mathrm{up}}}$ are the lower and 
upper mass limits respectively of the integration domain.

Solving this integral is the main task of stellar population synthesis modeling. Note that, although Eq. \ref{eq:muIMF} and the combination of Eq. \ref{eq:Ltot_sum} and Eq. \ref{eq:wi} would lead to the same 
mathematical expression, the interpretation of what is computed differs drastically. In the statistical case,
synthesis models compute the mean value of a probability distribution function (that can be split in components, or evolutionary phases), but, by virtue of the very probability concept, it is not needed that all the modeled phases should be present in an observed cluster. In the deterministic case, it is mandatory that all the considered phases are present (by construction), even if they correspond to an unphysical fractional number of stars (i.e fractional amounts). So, this statistical formulation includes by its very nature, the possible sampling effects in real stellar populations. As an important point, it is necessary to note that  sampling effects are not only a characteristic of the system under study, but also a characteristic of the observation (a narrow slit observation of a given system will include in the observation a lower number of stars than an observation with a broader slit).

Additionally, Eq. \ref{eq:muIMF} also reveals some of the problems and sources of uncertainty we have been talking about:

\begin{enumerate}
\item The discontinuities in the derivative $({d\ell(m;t)}/{dm})^{-1}$ are directly related to fast evolutionary phases and discontinuities in the input tracks and isochrones.

\item The fact that the models results come from an integral, $ \int_{m^\mathrm{low}}^{m^{\mathrm{up}}} \ell \, \varphi_\mathrm{M}\, dm$, instead of a {\it sum} of evolutionary phases, $\sum_i  w_i \,\ell_i(t)$, implies the need of selecting actual representative evolutionary phases, and not
the tabulated isochrone points as they are directly computed (especially if they are representative of a maximum luminosity, like the tip of the AGB).

\end{enumerate}

\section{Conclusions}\label{sect:conclusions}

 In this review we have addressed the current status of evolutionary synthesis models. We have shown that there are several sources of uncertainties:
 
 \begin{itemize}
 \item With respect to the input ingredients, the main source of uncertainty are interpolations. Such interpolations should ideally follow a physically-based scheme, but such scheme is not always possible.
We have shown here the physical assumptions of current interpolation schemes for tracks and atmospheres.

We have shown that the most 
reliable ages for model use and comparison are the 
MS turn-off ages of the used tracks. If differences appear at such ages, they reflect differences in the numerical methods used in each code, which must be further investigated.
 
 \item With respect to the computations aspect of synthesis models, we have reviewed the two main methods: isochrone synthesis and FCT.
Since both methods produce different results, we concluded that
the stellar population synthesis method are not a reliable tool yet.
For the time being, only  those regions of the electromagnetic spectrum where both methods coincide should be relied on as realistic outputs.

\item Regarding the numerical computations performed by the models, we have shown that quantities like 
the SN rate (or the rate of any given evolutionary phase) is a useful quantity to 
address both numerical and unphysical assumptions included in the models.

\item We have also reviewed additional uncertainties related to the incompleteness of the input ingredients, and we have outlined the issue of inclusion of stellar rotation in synthesis models.

\item Finally, we have discussed the use of synthesis models in realistic cases. We stress again that the results of the models depend also on the number of stars covered by the observation, and not only the number of stars in the system under study. We show how to address these problems 
with a statistical interpretation of the models' results.
 
 \end{itemize}

The main conclusion of this work is that the current literature about population synthesis does not provide enough information to address the uncertainties involved. In this sense, uncertainties can only be addressed if synthesis models papers explain completely and carefully their input hypotheses, in particular the interpolation schemes and its justification in physical or mathematical terms. An extensive documentation of codes (as Gary Ferland has done with his photoionization code) would highly improve the understanding and uncertainty evaluation in the synthesis method.

\begin{acknowledgements}
We thank the organizers (especially David Valls-Gabaud) for giving us the opportunity to give this review.
We also thank lots of people that at some moment have contributed to the development of  synthesis models, and all the colleagues with whom we have discussing about the subject of the limitations of the models. They include Jos\'e Miguel Mas Hesse, Marat Gilfanov, Steve Shore, Juan Betancort, Enrique P\'erez, David Valls-Gabaud, Georges Meynet, Daniel Schaerer, Sandro Bressan, Scilla Degl'Innocenti, Enzo Brocato, Leo Girardi, Gabriella Raimondo, Rosa Amelia Gonz\'alez, Gustavo Bruzual, Alberto Buzzoni, Carlos Rodrigo, Luis Manuel Sarro, J\'esus M\'aiz Apellan\'\i z, Miguel Ch\'avez, Emanuele Bertone, Roberto Cid Fernandes, Grazyna Stasi{\'n}ka, Laerte Sodr\'e, and Elena and Roberto Terlevich among others. We also  
acknowledge Nuno, Carlos, Dario and Eva for providing experimental facts about the discreteness of real populations.

This work has been supported by the Spanish {\it Programa Nacional de Astronom\'\i a y Astrof\'\i sica} through the project AYA2004-02703. MC is supported by a {\it Ram\'on y Cajal} fellowship. VL is supported by a {\it CSIC-I3P} fellowship. 
\end{acknowledgements}

\bibliographystyle{apj}
\end{document}